\newif\iffull
\fullfalse

\iffull
	\documentclass[11pt]{article}
        \usepackage[hyphens]{url}
        \usepackage[breaklinks,colorlinks]{hyperref}
        \usepackage[usenames,dvipsnames]{xcolor}
        \usepackage{lastpage}

        \title{\textbf{Privacy-Preserving and Verifiable Machine Learning as a Service}}

\else 	

        \documentclass[conference]{IEEEtran}
        \IEEEoverridecommandlockouts

        \title{Privacy-Preserving Verifiable Neural Network Inference Service}
        
        \author{\IEEEauthorblockN{Arman Riasi}
        \IEEEauthorblockA{
        \textit{Virginia Tech}\\
        armanriasi@vt.edu}
        \and
        \IEEEauthorblockN{Jorge Guajardo}
        \IEEEauthorblockA{
        \textit{Robert Bosch LLC — RTC}\\
        jorge.guajardomerchan@us.bosch.com}
        \and
        \IEEEauthorblockN{Thang Hoang}
        \IEEEauthorblockA{
        \textit{Virginia Tech}\\
        thanghoang@vt.edu}
        }

\fi

\usepackage{amsopn}

\usepackage{amssymb}
\usepackage{endnotes,microtype,xspace,graphicx,fancyvrb,multirow}
\usepackage{booktabs}
\usepackage{array,underscore,relsize}
\usepackage[T1]{fontenc}
\usepackage{fancyhdr}
\usepackage{enumitem}
\usepackage[labelfont=bf,font=small,skip=5pt]{caption}
\usepackage{fp}
\usepackage{siunitx}
\usepackage{amsmath}
\usepackage{comment}
\usepackage{array}
\newcolumntype{R}{>{\raggedleft\arraybackslash}X}  
\usepackage{seqsplit}
\usepackage{balance}

\usepackage{array}
\usepackage{cellspace}
\usepackage{etoolbox}
\usepackage{multicol}
\usepackage{fancybox}
\usepackage{pbox}

\usepackage[noend]{algpseudocode}
\usepackage{algorithm}
\usepackage{algorithmicx}
\usepackage[normalem]{ulem}

 \usepackage{tikz}
 \usetikzlibrary{plotmarks}
 \usepackage{pgfplots}
 \usetikzlibrary{patterns}
 \pgfplotsset{compat=1.3}
 \usetikzlibrary{backgrounds}

\usepackage{subcaption}
\usepackage{tabularx}
\usepackage{textcomp}

\usepackage{comment}
\usepackage{threeparttable}

\usepackage{amsthm}
\newtheorem{theorem}{Theorem}
\usepackage{tabularx}
\usepackage{seqsplit}

\usepackage{siunitx}

\usepackage{hyperref}
\usepackage{cleveref}
\usepackage{amsmath}
\usepackage{amsthm}

\def\Snospace~{\S{}}

\theoremstyle{definition}

\newcommand{\squishlist}{
\begin{itemize}[noitemsep,nolistsep]
  \setlength{\itemsep}{-0pt}
}
\newcommand{\squishend}{
  \end{itemize}
}

\newcommand{\enumlist}{
\begin{enumerate}[noitemsep,nolistsep]
  \setlength{\itemsep}{-0pt}
}
\newcommand{\enumend}{
\end{enumerate}
}

\usepackage{xstring}
\newcommand{\PP}[1]{
\vspace{2px}
\noindent{\bf {#1}{.}}
}
\newcommand{\PC}[1]{
\vspace{2px}
\noindent{\bf \IfEndWith{#1}{:}{#1}{#1:}}
}
\newcommand{\IP}[1]{
\vspace{2px}
\noindent{\it \IfEndWith{#1}{.}{#1}{#1.}}
}
\newcommand{\IC}[1]{
\vspace{2px}
\noindent{\it \IfEndWith{#1}{:}{#1}{#1:}}
}

\newcommand{\boxbeg}{
\noindent
\vspace{0.3em}
\begin{tabular}{|l|}
\begin{minipage}{0.94\columnwidth}
\vspace{0.1em}
\noindent
}

\newcommand{\boxend}{
\vspace{0.1em}
\end{minipage}\\ 
\end{tabular}
\vspace{0.3em}
}

\newcommand{\vect}[1]{\ensuremath{\mathbf{#1}}}

\newlist{myalg}{enumerate}{3}
\setlist[myalg]{
	topsep = 0in,
	parsep = 0in,
	partopsep = 0in,
	itemsep = 0in,
	labelsep = 0.07in,
	labelindent = 0in,
}
\setlist[myalg,1]{leftmargin=0.15in,label=\arabic*.,ref=\arabic*}
\setlist[myalg,2]{leftmargin=0.20in,label=(\alph*),ref=(\alph*)}
\setlist[myalg,3]{leftmargin=0.15in,label=\roman*.,ref=\roman*}
\newtheoremstyle{myProtocolStyle}
{}
{}
{}
{}
{%
	\bfseries
}
{.}
{ }
{\thmname{#1}\thmnumber{ #2}\thmnote{ (#3)}}%
\theoremstyle{myProtocolStyle}

\newtheoremstyle{myDefinitionStyle}
{}
{}
{\normalfont}
{}
{\bfseries}
{.}
{ }
{\thmname{#1}\thmnumber{ #2}\thmnote{ (#3)}}%

\newtheoremstyle{myTheoremStyle}
{}
{}
{\itshape}
{}
{%
	\bfseries
}
{.}
{ }
{\thmname{#1}\thmnumber{ #2}\thmnote{ (#3)}}%

\theoremstyle{myDefinitionStyle}
\newtheorem{Definition}{Definition}

\theoremstyle{myTheoremStyle}

\usepackage{amsthm}

\hypersetup{breaklinks,colorlinks,citecolor=blue,linkcolor=blue}

\usepackage{enumitem}
\setlist[itemize]{leftmargin=*}
\setlist[enumerate]{leftmargin=*}

\newcommand{\sys}{\ensuremath{\mathsf{vPIN}}} 
\newcommand{\SecondBaseline}{Baseline (w/o CE)}
\newcommand{\FirstBaseline}{Baseline (w/CE)}
\newcommand{\secParam}{\ensuremath{\lambda}}

\newcommand{\getsRandom}{\ensuremath{\stackrel{\$}{\gets}}}

\newcommand{\ZZ}{\ensuremath{\mathbb{Z}}}
\newcommand{\GG}{\ensuremath{\mathbb{G}}}

\newcommand{\negFunc}{\ensuremath{\mathsf{negl}}}

\algdef{SE}[DOWHILE]{Do}{doWhile}{\algorithmicdo}[1]{\algorithmicwhile\ #1}%

\newcommand\arman[1]{\textcolor{purple}{#1}}
\newcommand{\newarman}[1]{\textcolor{black}{#1}}
\newcommand{\rarman}[1]{\textcolor{black}{#1}}
\newcommand{\rarmantwo}[1]{\textcolor{blue}{#1}}

\newcommand{\circled}[2][]{%
	\tikz[baseline=(char.base)]{%
		\node[shape = circle, draw, fill=red, color=red, inner sep = .2pt]
		(char) {\phantom{\ifblank{#1}{#2}{#1}}};%
		\node at (char.center) {\makebox[0pt][c]{\color{white}{#2}}};}}
\robustify{\circled}

\newcommand{\Adv}{\ensuremath{\mathcal{A}}}
\newcommand{\Chal}{\ensuremath{\mathcal{Q}}}

\usepackage{stmaryrd}
\newcommand{\HECipher}[1]{\ensuremath{{\llbracket #1 \rrbracket}}}
\newcommand{\scalMult}[1]{\cdot #1}

\newcolumntype{L}{>{\centering\arraybackslash}X}

\newcommand{\Ra}{\ensuremath{\stackrel{\$}{\leftarrow}{\xspace}}}

\newcommand{\tikzxmark}{%
\tikz[scale=0.16] {
    \draw[line width=0.7,line cap=round] (0,0) to [bend left=6] (1,1);
    \draw[line width=0.7,line cap=round] (0.2,0.95) to [bend right=3] (0.8,0.05);
}}
\newcommand{\tikzcmark}{%
\tikz[scale=0.16] {
    \draw[line width=0.7,line cap=round] (0.25,0) to [bend left=10] (1,1);
    \draw[line width=0.8,line cap=round] (0,0.35) to [bend right=1] (0.23,0);
}}

\newcommand{\rel}{\ensuremath{\mathcal{R}}}
\newcommand{\Prover}{\ensuremath{\mathcal{P}}}
\newcommand{\Verifier}{\ensuremath{\mathcal{V}}}

\newcommand{\pp}{\ensuremath{\mathsf{pp}}}
\newcommand{\Extor}{\ensuremath{\mathcal{E}}}
\newcommand{\Sim}{\ensuremath{\mathcal{S}}}
\newcommand{\Comm}{\ensuremath{\mathsf{Comm}}}
\newcommand{\cm}{\ensuremath{\mathsf{cm}}}

\newcommand{\cpzkp}{\ensuremath{\mathsf{CPS}}}
\newcommand{\prove}{\ensuremath{\mathsf{{Prov}}}}
\newcommand{\ver}{\ensuremath{\mathsf{{Ver}}}}
\newcommand{\aux}{\ensuremath{\mathsf{{aux}}}}

\newcommand{\HE}{\ensuremath{\mathsf{AHE}}}
\newcommand{\sk}{\ensuremath{\mathsf{sk}}}
\newcommand{\pk}{\ensuremath{\mathsf{pk}}}
\newcommand{\KeyGen}{\ensuremath{\mathsf{KeyGen}}}
\newcommand{\Enc}{\ensuremath{\mathsf{Enc}}}
\newcommand{\Dec}{\ensuremath{\mathsf{Dec}}}
\newcommand{\parse}{\ensuremath{\textbf{parse }}}

\newcommand{\idealFunc}{\ensuremath{\mathcal{F}}}

\newcommand{\relu}{\ensuremath{\mathsf{TReLU}}}

\newcommand{\Infer}{\ensuremath{\mathsf{Infer}}}

\newcommand{\Setup}{\ensuremath{\mathsf{Setup}}}

\newcommand{\mat}[1]{\mathbf{#1}}

\newcommand{\MPP}{\ensuremath{\mathsf{M_{pp}}}}
\newcommand{\NPP}{\ensuremath{\mathsf{N_{pp}}}}
\newcommand{\RPP}{\ensuremath{\mathsf{R_{pp}}}}
\newcommand{\CPP}{\ensuremath{\mathsf{C_{pp}}}}
\newcommand{\DPP}{\ensuremath{\mathsf{D_{pp}}}}

\begin{document}

        \maketitle
        
        \begin{abstract}

Machine learning has revolutionized data analysis and pattern recognition, but its resource-intensive training has limited accessibility. 
Machine Learning as a Service (MLaaS) simplifies this by enabling users to delegate their data samples to an MLaaS provider and obtain the inference result using a pre-trained model. 
\newarman{Despite its convenience, leveraging MLaaS poses significant privacy and reliability concerns to the client.}
Specifically, sensitive information from the client inquiry data can be leaked to an adversarial MLaaS provider.
\newarman{
Meanwhile, the lack of a verifiability guarantee can potentially result in biased inference results or even unfair payment issues.
}
\newarman{
While existing trustworthy machine learning techniques, such as those relying on verifiable computation or secure computation, offer solutions to privacy and reliability concerns, they fall short of simultaneously protecting the privacy of client data and providing provable inference verifiability.%
}%

\newarman{In this paper, we propose $\sys$, a privacy-preserving and verifiable CNN inference scheme that preserves privacy for client data samples while ensuring verifiability for the inference.}
\sys~makes use of partial homomorphic encryption and \newarman{commit-and-prove succinct non-interactive argument of knowledge} techniques to achieve desirable security properties.
\newarman{In \sys, we develop various optimization techniques}
to minimize the proving circuit for homomorphic inference evaluation thereby, improving the efficiency and performance of our technique.  
We fully implemented and evaluated our $\sys$ scheme on standard datasets (e.g., MNIST, CIFAR-10).
\newarman{
Our experimental results show that $\sys$ achieves high efficiency in terms of proving time, verification time, and proof size, while providing client data privacy guarantees and provable verifiability.
}

\end{abstract}

                

	\section{Introduction}%
Machine learning (ML) has revolutionized the way computers interact with data, allowing them to acquire knowledge from data on their own, without the limitations of explicit programming. 
Deep learning, especially Convolutional Neural Networks (CNNs), has revolutionized the field of visual data analysis, enabling applications such as image classification and object recognition. 
\newarman{
However, addressing the extensive data demands of model training, computational resource constraints, and implementation expertise has led to the rise of Machine Learning as a Service (MLaaS). MLaaS is a cloud-based platform that provides access to powerful ML-assisted services (e.g., analysis, inference, prediction). Typically, MLaaS providers such as Microsoft Azure \cite{microsoftAzure}, Google Cloud AI Platform \cite{google}, Amazon AWS \cite{amazon}, Face++ \cite{faceplusplus}, and Clarifai \cite{clarifai} operate on a pay-as-you-go model, where clients pay based on their resource usage for inference.
}

Despite its benefits, 
delegating ML tasks to a server raises some privacy concerns. 
An adversarial server 
can misuse the private information of the client data. 
Additionally, 
the lack of a mechanism to verify ML operations performed by the server 
introduces integrity and trustworthiness issues. 
This becomes essential since an adversarial server may process the client request arbitrarily without relying on a dependable ML model. 
\newarman{Moreover, there is a risk of unfair payment, where the server may charge the client more than its actual resource consumption.
Consequently, the outcome of the delegated ML processing tasks can
be untrustworthy.
}

\newarman{
To address the aforementioned privacy and trustworthiness issues in ML, several research directions have been suggested. 
One line of research focuses on verifiable Machine Learning (verifiable ML), 
which aims to ensure the verifiability of ML computations and address concerns regarding unfair payment by requiring the server to provide mathematical proof of correct computations.
Some studies, such as Safetynets \cite{DNN1} and VeriML \cite{vML1}, investigate Verifiable Computation techniques (e.g., \cite{vc1, vc2}) that enable the client to effectively verify the correctness of computations performed by an MLaaS server. Moreover, some verifiable ML research incorporates zero knowledge property into proofs (e.g., \cite{cryptoProof1, cryptoProof2, cryptoProof3, cryptoProof4, spartan, cryptoProof5, bulletproofs}) to improve server model privacy. 
While these studies 
effectively address computation integrity and 
overcharging concerns, 
the main limitation of existing verifiable ML schemes is that 
they do not offer privacy to the client data, 
where the client has to send their sample (in plaintext) 
to the server for inference service. 
}

The other research line focuses on Privacy-Preserving ML (PPML), in which secure computation techniques such as Homomorphic Encryption (HE) \cite{HomomorphicEncryption} and/or Multi-party Computation (MPC) \cite{MPC} have been used, to protect the privacy of both client data and server model parameters during the ML evaluation.
However, these techniques may not be suitable for MLaaS applications.
\newarman{Specifically, HE-based PPML approaches \cite{LR3, infer4, infer6, infer1, infer2} do not offer computation verifiability. Consequently, the server can employ low-quality model parameters to process computation on encrypted client data, thereby making results unreliable.}
\newarman{Meanwhile, MPC-based approaches \cite{linearReg3, LRswift, ABY, kmean1} require distributed systems with multiple non-colluding computationally resourceful entities, which may significantly increase the deployment and operational costs.}

%

While existing research attempts to address privacy and verifiability concerns in ML inference schemes, 
a critical gap persists in ensuring client data privacy within verifiable ML schemes.
Therefore, we raise the following question:

\newarman{\textit{Can we design a new privacy-preserving and verifiable ML inference scheme that not only preserves the privacy of the client data
but also guarantees computational and provable verifiability with efficient performance?
}}

\PP{Our Contributions}%
\newarman{
In this paper, we introduce $\sys$, a new privacy-preserving and verifiable ML inference scheme that allows the client to use remote ML inference services with data privacy and inference result integrity guarantees.
\sys~focuses on the standard CNN inference framework with multiple processing layers such as convolution, activation, pooling, and fully connected layers.
\sys~relies on two core building blocks including partial homomorphic encryption and commit-and-prove Succinct Non-interactive Argument of Knowledge (SNARK) to enable client privacy and server computation integrity, respectively.
We provide detailed gadgets to represent arithmetic constraints for ciphertext operations (\autoref{sec:gadgets}).
%
These gadgets are not only critical for proving CNN inference evaluation on the encrypted data but also can be found useful in 
other applications. 
%
%
$\sys$ addresses the challenges of proving complicated ML functions on the encrypted domains by incorporating new techniques
to reduce the circuit complexity including 
elliptic curve embedding and probabilistic matrix multiplication check.
As a result, $\sys$ achieves markedly higher efficiency compared to the baseline approaches that hardcode the whole CNN inference computation on the encrypted data into the proving circuit.}

\newarman{We provide a formal security definition for 
privacy-preserving and verifiable ML inference with client data privacy and verifiability, 
and prove that \sys~satisfies all the security properties.
We fully implemented and evaluated our $\sys$ scheme, assessing its performance on real dataset and comparing it with the baseline setting in terms of proving time, verification time, and proof size. 
%
The source code for our implementation can be found at}

\hspace{4.7em}\url{https://github.com/vt-asaplab/vPIN}

\PP{Remark}%
\newarman{
In this work,
we only focus on user privacy and verifiability in place of server privacy.
It is a challenging task to enable privacy for both the user and server simultaneously.
Adding the zero-knowledge property to SNARK can only protect the server's model privacy in the proof; however, the model can still be leaked from the plain inference outputs (e.g., via model extraction/stealing attacks 
\cite{stealing1, stealing2, stealing3, stealing4, stealing5}). 
Therefore, we defer the investigation that can fully protect the server model privacy as our future work.
}

\PP{Applications Scenarios}%
\newarman{
Our proposed scheme 
can be useful in some applications in
medical diagnosis and financial forecasting \cite{MLaaSRisk1, MLaaSRisk2, MLaaSRisk3, MLaaSRisk4}.
For instance, 
in medical diagnosis,
healthcare providers 
might use MLaaS inference 
to analyze private client medical data, 
such as lung images, 
for infection detection.
However, 
incorrect model predictions 
could have a serious impact, 
especially if the cloud provider 
fails to use the actual model.
Additionally, 
outsourcing client medical data 
could potentially pose some complications.
To address them, 
\sys~allows healthcare providers 
to encrypt the image 
before sharing them with the MLaaS provider.
Moreover, 
using our scheme, 
the healthcare provider 
can verify that the correct model 
is used to process the image and 
ensure charge accuracy 
based on resource usage, 
thus preventing overbilling.
}

\newarman{
In financial forecasting, 
businesses may use MLaaS inference 
to analyze datasets and 
make decisions about 
investments, market trends, and risk management. 
Like medical diagnosis, 
these datasets contain sensitive business information 
that cannot be shared with an MLaaS provider.
Furthermore, 
ensuring the correct model parameters are used 
for prediction in financial forecasting is crucial, 
as it can have real-world implications 
(e.g., financial losses, missed opportunities).
Therefore, \sys~can address these concerns by processing financial data in encrypted form and providing inference verifiability to ensure the correct model is employed.
}
%

	\section{Preliminaries}

\PP{Notation}
%
%
Let $\ZZ_n = \{0,1,2,\dots,n\}$ and $ \ZZ_n^* = \ZZ_n \setminus \{0\} $.
$ \secParam $ denotes security parameters, and $ \mathsf{negl(\cdot)} $ stands for the negligible function.
$x \Ra \ZZ_n$ indicates that $x$ is chosen randomly from the set $\ZZ_n$.
PPT refers to Probabilistic Polynomial Time.
$A \stackrel{c}{\approx} B$ indicates computational indistinguishability of two quantities $A$ and $B$.
Bold letters (like $\vect{a}$ and $\mat{A}$) denote vector and matrix, respectively.
$\vect{a}[i]$ represents the element $i$ in vector $\vect{a}$.
$\mat{A}[i,j]$ refers to the element at row $i$ and column $j$ in matrix $\mat{A}$.
$\mat{A}^T$ denotes the transpose of $\mat{A}$.
$\GG$ represents the cyclic group in which $\HECipher{1} \in \GG$ denotes a random generator of $\GG$.
The group element $\HECipher{x}$ in $\GG$ has a discrete logarithm of $x$ with base $\HECipher{1}$,
where $x \in \ZZ_q$ ($q$ is the order of $\GG$). 
Group operations are additive, thus 
$\HECipher{x} + \HECipher{y}$ equals $\HECipher{x+y} \in \GG$
and
scalar multiplication over group $\GG$ is represented as  $\delta \scalMult{\HECipher{x}} \in \GG$, where scalar $\delta \in \ZZ_q$.
%
%
%
We denote the convolution operation between two matrices as $\mat{C} = \mat{A} * {\mat{B}}$ where $\mat{C}[i,j] = \sum_{i'=0}^{k-1} \sum_{j'=0}^{k-1} \mat{A}[i \cdot \hat{s} + i',j \cdot \hat{s} +j'] \scalMult{\mat{B}[i',j']}$, $\mat{B}$ is a $k \times k$ matrix, and $\hat{s}$ is the stride size.

\subsection{Additive Homomorphic Encryption}
Additive Homomorphic Encryption ($ \HE $) 
\cite{exponentialElgamal} is a public-key encryption that permits plaintext messages to be encrypted in a way that their ciphertexts can be homomorphically evaluated.
An $ \HE $ scheme consists of the following algorithms.
\begin{itemize}
	\item $ (\pk,\sk) \gets \HE.\KeyGen(1^\secParam) $: Given a security parameter \secParam, it outputs a pair of public key and private key $(\pk,\sk)$.
	\item $ c \gets \HE.\Enc(\pk, m) $: Given a message $ m $ and a public key \pk, it outputs a ciphertext $ c $.
	\item $ m \gets \HE.\Dec(\sk, c) $: Given a ciphertext $ c $ and a private key $ \sk $, it outputs a plaintext message $ m $.
\end{itemize} 

\PP{Homomorphism} Let $m_1 \in \mathbb{Z}_n$, $m_2 \in \mathbb{Z}_n$ be plaintext messages, $\HE$ offers additive homomorphic properties as

\begin{itemize}

\item \underline{Ciphertext addition}: Let $\boxplus$ be the group addition over the AHE ciphertext domain.  We have that
\begin{small}
$$
\setlength{\abovedisplayskip}{3pt}
\hspace{-2mm} 
\HE.\Dec(\sk, \HE.\Enc(\pk,m_1) \boxplus \HE.\Enc(\pk,m_2)) = (m_1 + m_2)
\setlength{\belowdisplayskip}{3pt}
$$
\end{small}

\item \underline{Scalar multiplication}:
Let $\boxdot$ be the group scalar multiplication over the AHE ciphertext domain. We have that
\begin{small}
$$
\setlength{\abovedisplayskip}{3pt}
\HE.\Dec(\sk, {m_2} \boxdot \HE.\Enc(\pk, m_1)) = m_1 \cdot m_2
\setlength{\belowdisplayskip}{3pt}
$$
\end{small}
\end{itemize}
%

We present the correctness and IND-CPA security of the AHE scheme in Appendix \autoref{sec:correctnessAndINDCPA}. 
%
%

\subsection{\newarman{Commit-and-Prove Argument Systems}} \label{sec:pre:zk}%
%
%
\label{sec:pre:zk}%
\newarman{Commit-and-Prove SNARK (CP-SNARK) \cite{cryptoProof1}
permits a prover to commit to a witness $w$ in advance 
for an NP-statement $(\rel, \mathcal{L})$, 
where $\rel$ is an NP relation and 
$\mathcal{L}$ is the language of valid inputs. 
Given $x \in \mathcal{L}$, 
the prover proves the knowledge of  $w$ 
that satisfies the NP relation $\rel$, 
i.e., $(x, w) \in \rel$.
CP-SNARK is a tuple of PPT algorithms $\cpzkp = ( \Setup, \Comm, \prove,\ver)$ as follows.}

\begin{itemize}
        \item $ \pp \gets \cpzkp.\Setup(1^\secParam) $: Given a security parameter $\secParam$, it outputs public parameters $\pp$.

        \item $ \cm \gets \cpzkp.\Comm(\vect{w}, r, \pp) $: Given a witness $\vect{w}$ and randomness $r$, it outputs $\cm$ as a commitment to $\vect{w}$.

        \item $ \pi \gets \cpzkp.\prove(\vect{x}, \vect{w}, \pp) $: Given a statement $\vect{x}$ and a witness $\vect{w}$, it outputs a proof $\pi$ that shows $(\vect{x}, \vect{w}) \in \rel$.


        \item $ b \gets \cpzkp.\ver(\pi, \cm, \newarman{\vect{x}}, \pp) $: Given a proof $\pi$, a commitment $\cm$, \newarman{and a statement $\vect{x}$},
        it outputs $b$ where $b$ is 1 (accept) if $\pi$ is a valid proof for $(\vect{x}, \vect{w}) \in \rel$ and $\cm$ is a commitment to $\vect{w}$; otherwise, it outputs 0 (reject).
        
\end{itemize}

\begin{Definition} \label{def:sec:cpzkp}
	\newarman{CP-SNARK satisfies the following security properties.}

\begin{itemize}
	\item \emph{Completeness:} For any $ \secParam, r, (\vect{x}, \vect{w}) \in \rel $  
        s.t.
        $\pp \gets \cpzkp.\Setup(1^\secParam)$, 
        $ \cm \gets \cpzkp.\Comm(\vect{w}, r, \pp) $, and $\pi \gets \cpzkp.\prove(\vect{x}, \vect{w}, \pp)$,
        \setlength{\abovedisplayskip}{3pt}
        \setlength{\belowdisplayskip}{3pt}
	\begin{gather*}
		\Pr \left[
		\cpzkp.\ver(\pi,\cm, \newarman{\vect{x}}, \pp) = 1
		\right] = 1
	\end{gather*}
        
 	\item \emph{Knowledge soundness:} For any PPT $ \Prover^* $, and statement $\vect{x}$, there exists a PPT extractor $ \Extor $ that, with access to the entire execution process and the randomness of $ \Prover^* $, can extract a witness $ \vect{w} $ s.t. $\pp \gets \cpzkp.\Setup(1^\secParam), \cm \gets \cpzkp.\Comm(\vect{w}, r, \pp) $ for randomness $r$, $\pi^*  \gets \Prover^*(\vect{x},\pp), \vect{w} \gets \Extor^{\Prover^*}(\vect{x}, \pi^*,\pp)$, it holds that 
        \begin{gather*}
		\Pr \left[
		(\vect{x}, \vect{w}) \notin \rel  \land \cpzkp.\ver(\pi^*, \cm, \newarman{\vect{x}}, \pp) = 1
		\right] \leq \mathsf{negl}(\secParam)
	\end{gather*}

\end{itemize}

\end{Definition}

%

%

%

\subsection{Convolutional Neural Networks}
CNNs \cite{CNN1, CNN2} allow
learning hierarchical features from data for tasks such as image classification and recognition.
CNNs consist of multiple consecutive layers as follows. 
%
%

%
\PP{Convolutional Layer}%
It extracts features from data samples by computing dot products between overlapping sections of an $n \times m$ input samples $\mat{X}$ and a $k \times k$ filter $\mat{K}$ as $\mat{A} = \mat{X} * {\mat{K}}$ (2D convolution).
%
%

\PP{Activation Layer}%
It introduces non-linearity by applying non-linear functions on the ${n}^\prime \times {m}^\prime$ convolution output to enable the network to learn complex and non-linear patterns in the data.
In this paper, we focus on ReLU as the activation function, which is defined as $\mat{B}[i,j] = \max(0, \mat{A}[i,j])$ for $i \in [0,{n}^\prime]$, $j \in [0,{m}^\prime]$.

\PP{Pooling Layer}%
It reduces the spatial dimensions of the ${n}^\prime \times {m}^\prime$ activation layer output by applying functions such as average pooling or max pooling. 
In this paper, we use the average pooling such that ${\mat{C}}[i, j] = \frac{1}{\hat{k}^2} \sum_{i'=0}^{\hat{k}-1} \sum_{j'=0}^{\hat{k}-1} {\mat{B}}[i \cdot \hat{k} + i', j \cdot \hat{k} + j']$ with $\hat{k} \times \hat{k}$ window.

\PP{Fully Connected (FC) Layer}%
It combines high-level features extracted from earlier layers to establish the complex relationships between the different features.
We define the FC layer as 
$\vect{e}[i] = (\sum_{j=0}^{g} \mat{W}^T[i,j] \scalMult{\vect{d}[j]}) + \vect{b}[i]$ for $i \in [0,h]$, 
where
$g, h$ denote the FC layer input and output sizes, respectively,
$\mat{W}$ is the model weight parameters, 
$\vect{d}$ 
is the $\hat{n} \times \hat{m}$ pooling layer output flattened 
such that 
$\vect{d}[i \cdot \hat{n} + j] = {\mat{C}[i,j]}$ for $i \in [0,\hat{n}]$, $j \in [0,\hat{m}]$, and $\vect{b}$ is the bias vector.
%

	\section{Model} \label{sec:model}%
\begin{figure}
    \centering
    \includegraphics[width=0.49\textwidth]{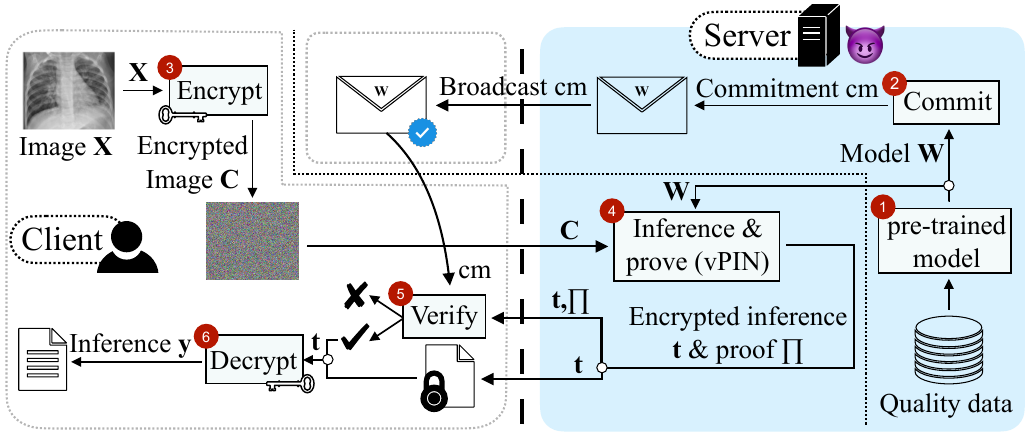}
    \caption{The overall architectural model of \sys.}
    \label{fig:sysmodel}
\end{figure}
\setlength{\floatsep}{0.1cm} 
\PP{System Model}%
Our $\sys$ system consists of two parties: the server and the client. 
The server holds well-trained ML model parameters $ \mat{W} \in \MPP$ where $\MPP$ denotes model space.
The client holds a private data sample $\mat{X} \in \NPP$, where $\NPP$ denotes message space. 
\newarman{
The server commits to the ML model $\mat{W}$ and allows a client to perform inference on her data sample $ \mat{X} $ using its committed model while preserving the privacy of the data sample, and guaranteeing computational correctness.}
\newarman{\autoref{fig:sysmodel} illustrates system workflow of \sys.}
Initially, 
we assume that the server owns an ML model  pre-trained from a high-quality data source (\circled[1]{\small 1}).
The server commits to this model beforehand (\circled[1]{\small 2}). 
In \sys, the client encrypts her data sample and sends it to the server (\circled[1]{\small 3}).
The server then performs the inference on the encrypted data and generates a proof of correct inference computations based on the committed model (\circled[1]{\small 4}).
Finally, the client verifies the proof (\circled[1]{\small 5}) and decrypts to obtain the inference result (\circled[1]{\small 6}).
%
%
\newarman{Formally speaking, our \sys~is a privacy-preserving and verifiable ML inference scheme as $\sys = (\Setup, \Comm, \Enc, \Infer, \Dec)$ as follows}.
\begin{itemize}
	\item $ \underline{(\sk, \pk, \pp) \gets \sys.\Setup(1^\secParam,l)} $: 
            Given a security parameter $\secParam$
            and an upper bound of the size of model parameter $l$, 
            it outputs a pair of public-private key $(\pk,\sk)$ 
            and public parameters $\pp$.
	
	\item $ \underline{\cm \gets \sys.\Comm(\mat{W},r,\pp)} $: 
            Given ML model parameters $\mat{W}$ $\in$ $\MPP$ 
            and a randomness $r \in \RPP$ 
            where $\RPP$ is the random space, 
            it outputs a commitment $ \cm \in \DPP$, where $\DPP$ is the commitment space.
	
	\item $ \underline{\mat{C} \gets \sys.\Enc(\pk, \mat{X}, \pp)} $: 
            Given a public key $ \pk $
            and a data sample $\mat{X}$ $\in$ $\NPP$, 
            it outputs a ciphertext $ \mat{C} \in \CPP$, where $\CPP$ denotes ciphertext space.
            
	\item $ \underline{(\pi,\vect{t}) \gets \sys.\Infer(\mat{W},\mat{C},\pp)} $: 
            Given ML model parameters $\mat{W}$ $\in$ $\MPP$
            and an encrypted data sample $\mat{C} \in \CPP$, 
            it outputs 
            an encrypted inference result $\vect{t} = \idealFunc(\mat{W},\mat{C}) \in \CPP$ (where $\idealFunc$ is the ideal ML inference functionality)
            and a proof $ \pi $. 
	
	\item $  \underline{\{\vect{y},\perp\} \gets \sys.\Dec(\sk, \vect{t},\cm, \pi, \pp)}$:  
            Given a private key $\sk$, 
            an encrypted inference $\vect{t} \in \CPP$, 
            a commitment $\cm \in \DPP$, 
            and a proof $\pi$, 
            it outputs a plaintext inference $ \vect{y} \in \NPP$ 
            if $ \pi $ is a valid proof for the inference computation; 
            otherwise it outputs $ \perp $.

\end{itemize}%

\PP{Threat Model}%
%
%
In our system, we assume the client to be honest. 
The server can behave maliciously to compromise client data and the integrity of the inference computation by employing arbitrary model parameters or an arbitrary data sample instead of the actual ones. 
In this paper, we aim to preserve the privacy of the client data sample against the server while ensuring the integrity of the inference computation.

\PP{Security Model}%
\newarman{
We define the security definition of our privacy-preserving and verifiable ML inference, which guarantees the privacy of data samples and the integrity of inferences as follows.
}

\begin{Definition} \label{def:sec:zkmlip}
	A privacy-preserving and verifiable ML satisfies the following security properties.
	\begin{itemize}
		\item \textbf{Completeness}.  
            For any 
            $\secParam, l$, 
		    {$ \mat{W} \in \MPP $}, $ \mat{X} \in \NPP$, $r \in \RPP$
        s.t.
		$ (\sk,\pk, \pp) \gets  \sys.\Setup(1^\secParam,l)$, 
		$\cm \gets \sys.\Comm(\mat{W},r,\pp)$,
		$\mat{C} \gets \sys.\Enc(\pk, \mat{X}, \pp)$, 
		$(\pi, \vect{t}) \gets \sys.\Infer(\mat{W},\mat{C},\pp)$,
            \setlength{\abovedisplayskip}{3pt}
            \setlength{\belowdisplayskip}{3pt}
            \begin{gather*}
			\Pr \left[
			\sys.\Dec(\sk,\vect{t},\cm,\pi,\pp) \ne \perp 
			\right] = 1
		\end{gather*}

        \item \textbf{Soundness}. 
        For any $\secParam, l$, 
        and $\mathrm{PPT}$ adversary $\mathcal{A}$, 
        it holds that
            \setlength{\abovedisplayskip}{3pt}  
            \setlength{\belowdisplayskip}{3pt}
		\begin{small}
			\begin{gather*}
				\Pr \left[
				\begin{aligned}
					(\sk,\pk,\pp) \gets  \sys.\Setup(1^\secParam,l) \\
					(\cm^*,\mat{W}^*, \mat{C}, \vect{t}^*,\pi^*,r) \gets \Adv(\pk, \pp) \\
                    \cm^* = \sys.\Comm(\mat{W}^*,r,\pp)\\
					\sys.\Dec(\sk, \vect{t}^*, \cm^*, \pi^*, \pp) \ne \perp   \\
					\idealFunc(\mat{W}^*,\mat{C}) \ne \vect{t}^* 			
				\end{aligned}
				\right] \leq \negFunc(\secParam)
			\end{gather*}
		\end{small}
    
	\item \textbf{Data sample and inference privacy.} 
        For any 
        $\secParam, l$, $ \mat{X} \in \NPP $ such that $(\sk,\pk,\pp) \gets \sys.\Setup(1^\secParam,l)$,   
        and PPT adversary $ \Adv $, there exists a simulator $ \Sim' $ such that 
        \setlength{\abovedisplayskip}{3pt}  
        \setlength{\belowdisplayskip}{3pt}
        \begin{small}
	\begin{gather*}
		\begin{aligned}
			& \Pr \left[
			\begin{array}{l}
				\Adv(\pk,\mat{C},\vect{t}, \pp)= 1
			\end{array}
			\middle \vert
			\begin{array}{r}
			\mat{C} \gets \sys.\Enc(\pk,\mat{X},\pp) \\
                    (\pi,\vect{t}) \gets \sys.\Infer(\mat{W}, \mat{C}, \pp)\\
                \end{array}
			\right] \\ \stackrel{c}{\approx}
			& \Pr \left[
			\begin{array}{l}
				\Adv(\pk,\mat{C}, \vect{t}, \pp)= 1
			\end{array} 
			\middle \vert 
			\begin{array}{r}
				(\vect{t}, \mat{C}) \gets \Sim'(\pk, |\mat{X}|,\pp) \\
			\end{array}
			\right]
		\end{aligned}
	\end{gather*}	
         \end{small}
	\end{itemize}
\end{Definition}

\PP{Out-of-Scope Attacks}%
%
%
In this paper, we assume the client to be honest and, therefore,
do not consider model extraction/stealing attacks \cite{stealing1, stealing2, stealing3, stealing4, stealing5} that aim to compromise the server model privacy by collecting query-response pairs from the target model. 
For dishonest clients, several countermeasures can be  considered such as model extraction detection \cite{watermarking1, stealing4, monitor, input1, output1}, limited queries \cite{input1}, query pattern recognition \cite{stealing4,monitor}, and differential privacy \cite{dp111, dp222}. 
We leave the investigation of how to apply these techniques under encryption as our future work.

	\section{Our Proposed Method}%
\sys~employs an AHE scheme based on an Elliptic Curve (EC) to encrypt user data. 
To prove the correctness of inference over encrypted data, we need to provide arithmetic constraints.
We first introduce necessary gadgets in \sys, which are intermediate arithmetic constraint systems for computation statements. We then present our \sys~construction in detail.

\subsection{Building Block Gadgets} \label{sec:gadgets}%
\newarman{
We introduce arithmetic constraints that are required for EC point addition and point multiplication.
}

\PP{Reverse Binarization Gadget}%
We \newarman{present} a reverse binarization gadget based on the binarization gadget \newarman{\cite{bulletproofs}}. 
Given a value $ a \in \ZZ_k $ and a vector $ \vect{v} \in \ZZ_2^n$, the reverse binarization gadget $\mathsf{RevBin}(a,\vect{v},n)$ can 
generate the arithmetic constraints to demonstrate that $ \vect{v} $ is a reverse binary representation of $ a $ as follows:
        \setlength{\abovedisplayskip}{7pt}  
        \setlength{\belowdisplayskip}{7pt}
	\begin{equation}
		\begin{cases}
			\vect{v}[i] \times \vect{v}[i] = \vect{v}[i] \text{ for } i \in [1,n]\\
			\sum_{i=1}^{n}\vect{v}[n-i] \cdot  2^{n-i-1} = a
		\end{cases} 
	\end{equation}

\PP{EC Point Addition Gadget}%
Given three EC points 
$A = (x_a, y_a) \in \ZZ_p^2$, 
$B = (x_b, y_b) \in \ZZ_p^2$, and
$C = (x_c, y_c) \in \ZZ_p^2$
on curve $E: y^2 = x^3 + {\alpha}x + \beta$ 
where ${\alpha}, {\beta} \in \ZZ_p$, we \newarman{present} a
point addition gadget $\mathsf{PtAdd}(C, A, B)$ 
to demonstrate the arithmetic constraints for $C = A + B$.
Let $\aux_1, \aux_2, \aux_3, \text{and } \aux_4 \in \ZZ_p$ be auxiliary witnesses. 
%
The set of arithmetic constraints for an ECC point addition is
    \setlength{\abovedisplayskip}{7pt}  
    \setlength{\belowdisplayskip}{7pt}
    \begin{equation}
	\begin{cases}
            \aux_1 = (x_b - x_a)^{-1} \mod{p}\\
            \aux_2 = (y_b - y_a) \cdot \aux_1 \mod{p}\\
            \aux_3 = \aux_2 \cdot \aux_2 \mod{p} \\
            x_c = \aux_3 - x_a - x_b \mod{p} \\
            \aux_4 = \aux_2 \cdot (x_a - x_c) \mod{p} \\
            y_c = \aux_4 - y_a \mod{p}
    \end{cases}	
 \end{equation}

\PP{EC Point Doubling Gadget}%
Given two EC points 
$A = (x_a, y_a)$ $\in \ZZ_p^2$ and
$D = (x_d, y_d) \in \ZZ_p^2$ 
on curve 
$E: y^2 = x^3 + {\alpha}x + \beta$ 
where ${\alpha}, {\beta} \in \ZZ_p$,
we \newarman{present} a 
point doubling gadget $\mathsf{PtDbl}(D, A)$ 
that demonstrates the arithmetic constraints for $D = 2A$.
Let $\aux_1, \aux_2, \aux_3, \aux_4, \text{and } \aux_5 \in \ZZ_p$ be auxiliary witnesses. 
%
The set of arithmetic constraints for an ECC point doubling is
    \setlength{\abovedisplayskip}{7pt}  
    \setlength{\belowdisplayskip}{7pt}
    \begin{equation}
	\begin{cases}
            \aux_1 = (2y_a)^{-1}\mod{p}\\
            \aux_2 = x_a \cdot x_a\mod{p} \\
            \aux_3 = (3\aux_2 + {\alpha}) \cdot \aux_1 \mod{p}\\
            \aux_4 = \aux_3 \cdot \aux_3\mod{p} \\
            x_d = \aux_4 - 2x_a \mod{p} \\
            \aux_5 = \aux_3 \cdot (x_a - x_d) \mod{p} \\
            y_d = \aux_5 - y_a \mod{p}
    \end{cases}	\label{eq:pointdbl}    
 \end{equation}

\PP{Point Multiplication Gadget}%
Given a scalar $w \in \ZZ_p$, 
two EC points 
$Q = (x_q, y_q) \in \ZZ_p^2$ 
and 
$P = (x_p, y_p) \in \ZZ_p^2$ 
on curve 
$E: y^2 = x^3 + {\alpha}x + \beta$ 
where ${\alpha}, {\beta} \in \ZZ_p$, 
we \newarman{present} point multiplication gadget $\mathsf{PtMul}(Q,w,P)$ that demonstrates the arithmetic constraints for 
$Q = w\scalMult{P}$.
%
%
\newarman{
Since we use the index-increasing approach of the \textit{double-and-add} algorithm, we employ the $\mathsf{RevBin}$ gadget to represent the scalar multiplier $w$ as a binary number in the reversed order, with the least significant bit (LSB) at index 0 and the most significant bit (MSB) at the last index of $\vect{v}$. 
Along with $\mathsf{RevBin}$, we also employ $\mathsf{PtAdd}$ and $\mathsf{PtDbl}$ gadgets to enable iterative point addition and doubling in the \textit{double-and-add} algorithm.
}
In that case, the arithmetic constraints are as follows:
    \setlength{\abovedisplayskip}{7pt}  
    \setlength{\belowdisplayskip}{7pt}
    \begin{equation}
	\begin{cases}
            \mathsf{RevBin}(w,\vect{v},n) \\

            A_0 = P \\
            B_0 = \infty\footnotemark \\
 
            \mathsf{PtAdd}(C_i, B_{i-1}, A_{i-1}) \text{ for } i \in [1,n] \\    
            \mathsf{PtDbl}(D_i, A_{i-1}) \text{ for } i \in [1,n] \\
            
            B_i = (\vect{v}[i] \scalMult{C_i}) + ((1-\vect{v}[i]) \scalMult{C_{i-1}})\mod{p} \text{ for } i \in [1,n]\\ 
            
            A_i = D_i \text{ for } i \in [1,n] \\ 
            
            Q =  B_{n} \\
            
    \end{cases}	\label{eq:pointMul} 
 \end{equation}\vspace{-2em}
\footnotetext{This can be achieved using an indicator for the point at infinity.}

\subsection{\newarman{Detailed $\sys$ Construction}} \label{sec:proposedmethod}

%
We are now ready to present \sys~in detail, including its setup, commitment, encryption, inference, and decryption algorithms.
Due to space constraints, we present the pseudocode of these algorithms in Appendix \autoref{sec:appendixDetailedProtocol}.

\subsubsection{Setup procedure}%
 
This procedure generates the necessary parameters to ensure the privacy of the client's data sample and the verifiability of inference computations. 
These include public parameters \pp~for the underlying CP-SNARK and a public-private key pair $(\sk,\pk)$ for the back-end AHE scheme.
To do so,
we first identify suitable cyclic groups (i.e., EC parameters) used in CP-SNARK and AHE.
Note that in \sys, 
the server performs homomorphic evaluations between its model parameter and the (encrypted) client's sample and creates a proof of correct homomorphic evaluation. 
That means the proof circuit for such homomorphic evaluations can be extremely large if the parameters are not selected properly.

For instance,
let curve $E_1$ be the cyclic group of CP-SNARK with base field $\ZZ_{n_1}$ and order ${q_1}$, and curve $E_2$ be the cyclic group of AHE with base field $\ZZ_{n_2}$ and order $\ZZ_{q_2}$.
%
%
Let $w \in \ZZ_{q_2} $ be the (plaintext) server model, and $\HECipher{C}_2 = (x_{c}, y_{c}) \in \ZZ_{n_2}^2$ 
be the AHE-encrypted client data sample.
Suppose $w = 2$ for simplicity, 
the set of arithmetic constraints under CP-SNARK based on (\ref{eq:pointdbl}) for the homomorphic evaluation of $ \HECipher{w\cdot C}_2$ is as follows:
\setlength{\abovedisplayskip}{7pt}  
\setlength{\belowdisplayskip}{7pt}
\begin{equation}
	\begin{cases}
        \aux_1 = ((2y_c)^{-1}\mod{n_2}) \mod q_1\\
        \aux_2 = (x_c \cdot x_c\mod{n_2}) \mod q_1 \\
        \aux_3 = ((3\aux_2 + {\alpha}) \cdot \aux_1 \mod{n_2}) \mod q_1\\
        \aux_4 = (\aux_3 \cdot \aux_3\mod{n_2}) \mod q_1 \\
        x_d = (\aux_4 - 2x_c \mod{n_2}) \mod q_1 \\
        \aux_5 = (\aux_3 \cdot (x_c - x_d) \mod{n_2}) \mod q_1 \\
        y_d = (\aux_5 - y_c \mod{n_2}) \mod q_1
    \end{cases}	
        \label{eq:costlyMod}    
 \end{equation}


\newarman{
We can see that each arithmetic constraint in (\ref{eq:costlyMod}) 
requires creating costly modular arithmetic circuits 
for proving modulo $n_2$. 
For instance, 
to prove the relation 
$\hat{\aux} = \hat{x} \cdot \hat{y}\mod{n_2}$ 
where $\hat{x},\hat{y} \in \ZZ_{n_2}$, 
we require 23 constraints
according to the most efficient modular multiplication gadget in \cite{verizexe}.
Thus, proving a \textit{single} point doubling or point addition involves 161 and 230 arithmetic constraints, respectively.
Moreover, 
when these gadgets are iteratively used to prove just \textit{one} point multiplication with a 128-bit scalar in (\ref{eq:pointMul}), 
they dramatically increase the number of arithmetic constraints to 76834.}

\newarman{
To avoid the need for creating costly arithmetic constraints for proving modulo $n_2$,
\sys~employes} two correlated curves $E_1$, $E_2$ for CP-SNARK and AHE, respectively, by implementing a so-called curve embedding technique \cite{CurveEmbedding1, CurveEmbedding2, zeestar}.
Specifically,
we find a custom curve for $E_2$ such that its base field size equals the order of $E_1$, i.e., $n_2 = q_1$.
\newarman{This allows the AHE homomorphic evaluation to be directly captured in the CP-SNARK proving circuits, eliminating the need for creating constraints for proving modulo operations.
As a result, the set of arithmetic constraints is reduced, leading to
more efficient proving computation, verification, and proof size.}
%

\subsubsection{Commitment}%
%
%
Given a CNN model $\mat{W}$ with three sets of parameters
(including convolutional filter $\mat{F}$, and weights $\mat{\hat{W}}$ and biases $\hat{\vect{b}}$ for FC layer), 
the server commits to $\mat{W}$ using the commitment protocol 
of the back-end CP-SNARK, and publishes the commitment \cm~to the client (see \autoref{fig:PVMLprotocol1} in Appendix \autoref{sec:appendixDetailedProtocol}).
%


%

%
%


\subsubsection{Encryption}%
In \sys, the client encrypts data sample $\mat{X}^\prime$ 
before sending it to the server for inference. 
As $\mat{X}^\prime$ can be real numbers in practice, 
$\sys$ makes use of fixed-point representation to encode $\mat{X}'$ as elements in $\ZZ_p$ for AHE encryption. 
For each value $\mat{X}^\prime[i,j] \in \mathbb{R}$, 
the client computes its two's complement fixed-point representation and encrypts it with AHE as $\HECipher{\mat{C}[i,j]}_2 \gets \HE.\Enc(\pk,\mat{X}[i,j])$, where $\mat{X}[i,j]= \mat{X}^\prime[i,j] \cdot 2^f$, and $f$ is a system parameter indicating the number of fractional bits (e.g., $f = 16$) in the fixed-point representation (\autoref{fig:PVMLprotocol1}).
To this end, the client sends the encrypted sample $\HECipher{\mat{C}}_2$ to the server for inference evaluation.

\subsubsection{Inference}%
We present how the server evaluates the CNN inference on the encrypted sample and generates the corresponding proof. 
%
Intuitively,
\sys~realizes the privacy-preserving CNN inference via additive homomorphic properties of AHE, while the proof is generated by creating arithmetic circuits with EC gadgets introduced in \autoref{sec:gadgets}.
\sys~follows the standard CNN inference computation, 
which consists of a sequence of operations in the convolutional layer, average pooling, FC layers, and activation layers as follows. 

\PP{Convolutional Layer}%
The server first computes $\HECipher{\mat{A}}_2 = \HECipher{\mat{F} * \mat{C}}_2$, the convolution between 
its  convolution filter $\mat{F}$  of size $k\times k$
and the client sample $\HECipher{\mat{C}}_2$ of size $n \times m$ with a stride size $\hat{s}$ by evaluating the homomorphic dot products 
as
\begin{equation}\label{eq:HE:Conv}
\HECipher{\mat{A}[i,j]}_2 = \sum_{i'=0}^{k-1} \sum_{j'=0}^{k-1} \mat{F}[i',j'] \scalMult{\HECipher{\mat{C}[i \cdot \hat{s} +i',j \cdot \hat{s}+j']}_2} 
\end{equation}
for $i \in [0, n']$ and $j \in [0,m']$, where $n^\prime = (((n - k)/\hat{s}) + 1)$ and $m^\prime = (((m - k)/\hat{s}) + 1)$. 

\PP{Activation Layer}%
In the standard CNN,
after convolution, the next step is to apply an activation function.
Since the activation function is generally non-linear, 
it cannot be evaluated with AHE.
\newarman{Therefore, in \sys,
we delegate the evaluation of the activation function to the client by having the server transmit the encrypted convolution evaluation.}
Another important aspect of having such a client-server interaction is to handle the restricted plaintext size of AHE, particularly the exponential ElGamal. 
Specifically, the convolution output can be up to 35 bits due to dot product operations between the server model and the client sample.
Once this output is used for next-layer evaluation,
the result can grow up to a size (e.g., >80 bits) that cannot be decrypted by the discrete log algorithm. 
To handle this issue, 
in \sys, after each layer evaluation, the server will send the result to the client so that she can perform a truncation to keep the plaintext size within the decryptable range.
In the activation layer, the interaction works as follows.

Let $\HECipher{\mat{A}}_2 = \HECipher{\mat{F} * \mat{C}}_2$ be the encrypted convolution evaluation.
In \sys, 
for simplicity, 
we use ReLU as the activation function (i.e., $a' \gets \mathsf{ReLU}(a)$, where $a' = \max(0,a)$ for each $a \in \mat{A}$).
However, other activation functions such as Sigmoid, Tanh, Swish, and Mish can also be employed effectively.
To evaluate ReLU, the server sends $\HECipher{\mat{A}}_2$ to the client. 
For each $\HECipher{a}_2 \in \HECipher{\mat{A}}_2$, the client first 
decrypts it to obtain the value ${a}$ and then evaluates the ReLU function by sending back to the server 
either 
$\HECipher{a'}_2 = \HECipher{0}_2$ if $a \leq 0$ or
$\HECipher{a'}_2 = \HECipher{{{a}}^*}_2$ if $a > 0$, 
where ${{a}^*} = {{a}}/2^{\zeta}$ is the $f$-bit truncated value of ${a}$, and $\zeta$ is the number of truncated bits.
To this end, the server obtains the encrypted ReLU evaluation.
%
We represent our $\relu$ activation function in Appendix \autoref{sec:appendixDetailedProtocol}, \autoref{fig:Relu}.

\PP{Pooling Layer}%
Let $\HECipher{\mat{A}'}_2$ be the encrypted output of the activation.
The next step is to apply a pooling layer on $\HECipher{\mat{A}'}_2$.
Since $\HECipher{\mat{A}'}_2$ is an AHE cipher, 
we opt for a linear pooling function (i.e., average pooling) so that 
it can be evaluated homomorphically by the server.
The server computes the average pooling with the $\hat{k} \times \hat{k}$ window as 
\begin{equation}\label{eq:HE:avgpool}
\HECipher{\mat{B}[i,j]}_2 = \frac{1}{\hat{k}^2} \sum_{i'=0}^{\hat{k}-1} \sum_{j'=0}^{\hat{k}-1} \HECipher{\mat{A}^\prime[i \cdot \hat{k} + i', j \cdot \hat{k} + j']}_2
\end{equation}
for $i \in [0, n']$ and $j \in [0,m']$. 
%

%
In each step of the average pooling computation, the server slides a $\hat{k} \times \hat{k}$ window across the input $\HECipher{\mat{A}'}_2$. 
At each step, the server computes the homomorphic sum of the window's elements and obtains the average by multiplying it with the fixed-point representation of $1/\hat{k}^2$. 

\PP{FC Layer}%
Let $\HECipher{\vect{d}}$ be the output of average pooling flattened by $\HECipher{\vect{d}[i\cdot \hat{n} + j]}_2 = \HECipher{\mat{B}[i,j]}_2$ for $i \in [0, \hat{n}]$ and $j \in [0,\hat{m}]$, where $\hat{n} \times \hat{m}$ denotes $\HECipher{\mat{B}}_2$ dimensions. 
However, since the $\vect{\hat{b}}$ values are real numbers, the server first computes the fixed-point representation for each value $\vect{\hat{b}}[i]$ and encrypts it with AHE as $\HECipher{\vect{b}[i]}_2 \gets \HE.\Enc(\pk,\vect{\hat{b}}[i])$ (lines 28-29).
Next, the server computes the FC layer with an input size of $g$ and an output size of $h$. Hence, the server performs homomorphic additive sum on these encrypted bias elements $\HECipher{\vect{b}}_2$ along with the result obtained from the point multiplication, resulting in the inference result $\HECipher{\vect{t}}_2$ as
\begin{equation}\label{eq:HE:fclayer}
\HECipher{\vect{t}[i]}_2 = (\sum_{j=0}^{g} \hat{\mat{W}}^T[i,j] \scalMult{\HECipher{\vect{d}[j]}_2}) + \HECipher{\vect{b}[i]}_2 
\end{equation}
for $i \in [0,h]$. 

If there are multiple FC layers, the subsequent step between each FC layer is to compute the $\relu$ function 
where the operation is similar to the first activation layer. 

\PP{Proving Convolutional Layer Computation}%
As the client sample is encrypted, the convolutional layer 
comprises two main operations including EC point multiplication and EC point addition.
Thus, to prove the convolution layer, we can employ $\mathsf{PtAdd}$ and $\mathsf{PtMul}$ gadgets to generate the corresponding arithmetic constraints.
However, since (\ref{eq:HE:Conv}) incurs $O(n^2 \cdot k^2)$ EC point multiplications and 
$O(n^2 \cdot k^2)$ EC point additions 
to compute $\HECipher{\mat{A}}_2$ (assume $m =n$ and $\hat{s}$ is small constant), 
simply proving each point operation with an EC gadget will result in a significant cost.

To reduce the overhead, the high-level idea is to flatten components in the convolution layer and then apply the random linear combination technique \cite{Freivalds}.
Specifically,
each convolution iteration by (\ref{eq:HE:Conv}) 
applies a sliding window of size $k\times k$ to $\HECipher{\mat{C}}_2$ to extract $k^2$ elements in $\HECipher{\mat{C}}_2$ for dot product computation with the filter $\mat{F}$.
First, we flatten each of such windows to a vector of size $k^2$, and then form a matrix $\HECipher{\hat{\mat{C}}}_2$ that contains all the flattened vectors. 
Specifically, let $n^\prime = (((n - k)/\hat{s}) + 1)$, $m^\prime = (((m - k)/\hat{s}) + 1)$ be convolution output size, 
$\HECipher{\hat{\mat{C}}}_2$ is a matrix of size $n'\cdot m' \times k^2$ such that, for each
$i \in [0,n']$, 
$j \in [0,m']$,
$i' \in [0,k]$, 
$j' \in [0,k]$,
\begin{small}
        \begin{equation*}
                \HECipher{\mat{\hat{C}}[i \cdot m' + j, i' \cdot k + j']}_2 = \HECipher{\mat{C}[i \cdot \hat{s} + i', j \cdot \hat{s} + j']}_2 
                \label{eq:chatformula}
        \end{equation*}
\end{small}
Next, we flatten the convolution filter matrix $\mat{F}$ to a vector $\vect{\hat{f}}$, where $\vect{\hat{f}}[i.k + j]  = \mat{F}[i,j]$ for $i \in [0,k]$, $j \in [0,k]$.
Similarly, we flatten the convolution output as 
$\HECipher{\vect{\hat{a}}[i.m' + j]}_2 = \HECipher{\mat{A}[i,j]}_2$ for $i \in [0,n']$, $j \in [0,m']$ and then apply the random linear combination with a random challenge $\gamma$ (chosen by the verifier) to all the flattened components.
Specifically, 
(5) holds for all $i \in [0,n'] \text{ and } j \in [0,m']$ is equivalent iff the following equation holds
\begin{equation}
        \sum_{i'=1}^{n' \cdot m'} \gamma^{i'} \cdot \HECipher{\vect{\hat{a}}[i']}_2 = \sum_{j'=1}^{k^2} \biggl( \sum_{i'=1}^{n' \cdot m'} \gamma^{i'} \cdot \HECipher{\mat{\hat{C}}[i']}_2 \biggr)^T \cdot \vect{\hat{f}}[j']
        \label{eq:convRandomLinearCombination1}
\end{equation}
We can see that the number of EC point multiplications and EC point additions in 
(\ref{eq:convRandomLinearCombination1}) is reduced from $O(n^2 \cdot k^2)$ to $O(k^2)$.

\PP{Proving Average Pooling}%
In \sys, the average pooling only computes
the homomorphic sum of encrypted values followed by multiplying with a public window size parameter (i.e., $\hat{k}^{-2} \in \ZZ_{q_1}$).
Therefore, it suffices to prove this layer by employing $\mathsf{PtAdd}$ gadgets for arithmetic constraint representation in \eqref{eq:HE:avgpool}.
%
%
%
%
%
%
%

\PP{Proving FC Layer}%
This layer incurs EC point additions and multiplications. 
Therefore, 
we can use $\mathsf{PtAdd}$ and $\mathsf{PtMul}$ gadgets to represent corresponding arithmetic constraints. 
Nevertheless, 
simply using EC gadget to prove 
\eqref{eq:HE:fclayer}
will be costly since it incurs
$O(g \cdot h)$ EC point additions and multiplications to compute $\HECipher{\vect{t}}_2$, where $g, h$ indicates the size of the FC layer input and output, respectively.

To reduce the number of constraints in proving the FC layer,
we employ the random linear combination similar to the proof of the convolution layer. 
However, unlike the convolution layer, 
we do not need to flatten the weight matrix $\hat{\mat{W}}$ in the FC layer before applying the random linear combination. 
This is because the convolution layer involves multiple matrix multiplications between the filter $\vect{\hat{f}}$ and the (small) windowed inputs $\HECipher{\mat{C}}_2$. 
To perform an efficient random linear combination on the convolution, each window is flattened and formed into a big matrix $\HECipher{\hat{\mat{C}}}_2$, enabling a single matrix multiplication between the filter $\vect{\hat{f}}$ and $\HECipher{\hat{\mat{C}}}_2$. 
Since the FC layer incurs only a single matrix multiplication between the weight  $\mat{\hat{W}}$ and the output of the average pooling $\HECipher{\vect{d}}_2$, the flattening is not required.
%
%
%
%
Let $\gamma'$ be a random challenge chosen by the verifier.
The FC computation in \eqref{eq:HE:fclayer} holds if the following equation holds 
%
%
\begin{equation}
        \sum_{i'=1}^{h} {\gamma'}^{i'} \cdot \HECipher{\vect{t}[i']}_2 = \sum_{j'=1}^{g} \biggl( \sum_{i'=1}^{h} {\gamma'}^{i'} \cdot \mat{\hat{W}}^T[i'] \biggr) \cdot \HECipher{\vect{d}[j']}_2
        \label{eq:fclayerrandomlinearcombination}
\end{equation}
%
%
%
%
In \eqref{eq:fclayerrandomlinearcombination}, the number of constraints for EC point multiplications and additions is reduced from $O(g \cdot h)$ to $O(g)$ and $O(g+h)$, respectively.
We present the pseudocode of the \sys~inference procedure in \autoref{fig:PVMLprotocol2} in Appendix \autoref{sec:appendixDetailedProtocol}.

\PP{Remark}%
In our $\sys$ scheme, the server computes the convolutional, average pooling, and FC layers, while the client performs activation. Thus, the server only proves the correctness of the convolutional, average pooling, and FC layers.
%


\subsubsection{Decryption}%
In this procedure, the client receives the encrypted inference and the proof of correct homomorphic evaluation of convolution, pooling, and FC layers with respect to the server' committed model.
The client verifies the proof, and, 
if the proof is valid, 
she decrypts the encrypted inference and obtains the final result  (see \autoref{fig:PVMLprotocol3} in Appendix \autoref{sec:appendixDetailedProtocol}). 
	\section{Analysis}  \label{sec:analysis}%

\begin{table*}[h]
\small 
\centering 
\begin{threeparttable}
\caption{Complexity of $\sys$ (vs. \newarman{hardcoding baseline}).}
\label{tab:complexity}
\renewcommand{\arraystretch}{0.8} 
\begin{tabularx}{\linewidth}{@{} l X@{\hspace{-0.6em}} X@{\hspace{-0.6em}} c X X @{}}
\toprule
& \multicolumn{2}{c}{$\sys$} & \phantom{abc}& \multicolumn{2}{c}{\textbf{Baseline}}\\
\cmidrule{2-3} \cmidrule{5-6}
& \# point mult & \# point add && \# point mult & \# point add\\ %
\midrule
Conv. & $O(k^2)$ & $O(k^2)$ && $O(n^2 \cdot k^2)$ & $O(n^2 \cdot k^2)$\\ 
Avg Pool & 0 & $O((n+k)^2)$ && 0 & $O((n+k)^2)$\\
FC & $O(g)$ & $O(g+h)$ && $O(g \cdot h)$ & $O(g \cdot h)$\\
\midrule
Total & $O(k^2 + g)$ & $O((n+k)^2 + g + h)$ && $O((n^2 \cdot k^2) + (g \cdot h))$ & $O((n^2 \cdot k^2) + (n+k)^2 + (g \cdot h))$\\
\bottomrule 
\end{tabularx}
\end{threeparttable}
\end{table*}

\PP{Complexity}%
Let $k \times k$ be the filter dimension in the convolutional layer, $n \times m$ be the dimension of the client data sample  (for simplicity, we assume $m = n$),
and $\hat{s}$ be the stride size of the convolution window.
Let $\hat{k} \times \hat{k}$ be the window dimension in the average pooling layer
and $g,h$ be the input and output size of the FC layer.
In the convolutional layer, 
\sys~incurs $k^2 = O(k^2)$ point multiplications and $k^2-1 = O(k^2)$ point additions to prove the convolution layer (given $\hat{s}$ as small constant), compared with 
$ O(n^2 \cdot k^2)$ 
point multiplications and  
$O(n^2 \cdot k^2)$
point additions in the baseline approach that hardcodes the convolution to the proving circuit.
\sys~incurs $(\hat{k}^2 - 1)(((n - k)/\hat{s}) + 1)^2 /\hat{k}^2 = O((n+k)^2)$ point additions to prove pooling layer (with the average pooling).
In the FC layer, \sys~requires $g = O(g)$ point multiplications and $(g - 1) + h  = O(g+h)$ point additions compared with  $O(g \cdot h)$ point multiplications and additions in the baseline.
In overall, the complexity of~\sys~is $O(k^2 + g)$ point multiplications and $O((n+k)^2 + g + h)$ point additions (\autoref{tab:complexity}).

\PP{Security} \newarman{We state the security of \sys~as follows.}
\begin{theorem}\label{thm:Theorem1}
Assuming AHE is IND-CPA secure by \autoref{def:sec:indcpa} and the back-end CP-SNARK is secure by \autoref{def:sec:cpzkp}, 
$\sys$ 
is a  privacy-preserving and verifiable ML inference by \autoref{def:sec:zkmlip}.
\end{theorem}

\begin{proof}
    \newarman{See Appendix \autoref{sec:securityProof}.}
\end{proof}

	\section{Experimental Evaluation}

\subsection{Implementation}%
We fully implemented our $\sys$ scheme in Python and Rust with $\approx$ 5200 lines of code.
We employed the \texttt{python-ecdsa} library \cite{lib:pythonecdsa} to generate cryptographic keys based on the EC.
We used the standard Python \texttt{socket} for client-server communication and employed Python's \texttt{hmac} and \texttt{hashlib} cryptographic libraries to implement pseudo-random functions (PRFs) for random linear combinations.
We implemented the CNN inference phase for several CNN networks, including LeNet \cite{lenet}, from scratch to obtain all witnesses for generating the proofs and committing to auxiliary witnesses.
We also implemented the baby-step giant-step algorithm \cite{bsgs} for computing the EC discrete logarithm in our decryption phase.
We used the Rust \texttt{libspartan} library \cite{lib:spartan} to convert our arithmetic constraints into a rank-1 constraint system (R1CS) and implement our underlying CP-SNARK scheme involving witnesses commitment, proof generation, and verification.
Specifically, we used \texttt{Spartan\textsubscript{DL}} variant, 
which internally leveraged \texttt{Hyrax} polynomial commitment \cite{hyrax}, 
and \texttt{curve25519-dalek} \cite{lib:curveDalek} for EC operations.
\newarman{Our implementation is available at }\url{https://github.com/vt-asaplab/vPIN}.

\subsection{Configuration}%
\PP{Hardware}%
For the client, we used a local Macbook Pro 2021 with a $3.2$ GHz M1 Pro CPU and $16$ GB RAM. For the server, we used a server machine with a 48-core Intel CPU (Intel(R) Xeon(R) Platinum 8360Y CPU @ 2.40GHz) and \rarman{256 GB of RAM}.
We used thread-level parallelization in \texttt{libspartan} to speed up the proving/verification time.
Therefore the experimental results reported here are for multi-thread execution, as specified in their respective sections.

\PP{System Parameters}%
For the curve $E_1$, we used the standard curve \texttt{curve25519-dalek} in \texttt{libspartan} with 128-bit security.
$E_1$ used $y^{2}=x^{3}+486662x^{2}+x$
with based field $2^{255} - 19$ 
and prime order ${2^{252}+ \seqsplit{27742317777372353535851937790883648493}}$.
We generate the parameters
for the curve $E_2$ using the ECB toolkit \cite{app:ECB} to ensure the base field of $E_2$ matches the prime order of $E_1$.
$E_2$ used 
$y^2 = x^3 + \alpha_2 x + \beta_2$\footnote{The base-64 encoding of $\alpha_2$ is \texttt{\seqsplit{B7gQfOmTdkBeTbfbAw9XyzpLLxAP9Z9EjiYqP2UyG50=}}
\\The base-64 encoding of $\beta_2$ is \texttt{\seqsplit{CAi4LFqrcPqSXatviSmVBGR+j78B7HY4+UDsbkTKU1Y=}}}
with based field $2^{252}+\seqsplit{27742317777372353535851937790883648493}$
and prime order 
$2^{252} - \seqsplit{124614587218531604318505012771651942947}$.


\PP{Dataset}
We evaluated the performance of our proposed method on the MNIST dataset. 
We resized the MNIST dataset images from 28$\times$28 to 32$\times$32 to be suitable to the selected CNN parameter  networks.

\PP{CNN Network Parameters}%
We selected different CNNs labeled as networks A, B, C, D, and E (with the increase of model size and number of parameters), each featuring a single convolutional layer, a pooling layer, two FC layers, and two activation layers. 
All networks use a $3 \times 3$ filter size for the convolutional layer. 
Networks C, D, and E employ a $2 \times 2$ window size for the pooling layer, while networks A and B use a $4 \times 4$ window size. 
All networks consist of two FC layers with the number of neural nodes ranging from 10 to 256 per layer. 
For networks A and B, the first FC layer has 64 nodes, and ends with 16 and 32 nodes in the output, respectively. 
For networks C, D, and E, the first FC layer starts with 256  nodes and outputs with 16, 32, and 64 nodes, respectively. 
Within the second FC layer, networks A and B have 16 and 32 nodes, respectively, whereas networks C, D, and E feature 16,32, and 64 nodes, respectively. 
All networks have 10 output nodes.

We also assessed the performance of $\sys$ with the LeNet network \cite{lenet}. The LeNet model has a total of 61,706 parameters consisting of 3 convolutional layers, 2 average pooling layers, and 2 FC layers. 
Specifically, there are 6 feature maps in the first convolutional layer and first average pooling, 16 feature maps in the second convolutional layer and second average pooling, and 120 feature maps in the third convolutional layer.

\begin{figure*}[!t] 
	\centering
		\resizebox{1.01\textwidth}{!}{
			\begin{subfigure}{0.3\textwidth}
%
%
\definecolor{A}{HTML}{e6194B}%
\definecolor{B}{HTML}{f58231}%
\definecolor{C}{HTML}{4363d8}%
\definecolor{D}{HTML}{911eb4}%
\definecolor{E}{HTML}{3cb44b}%
\definecolor{F}{rgb}{0.92900,0.69400,0.12500}%
\definecolor{G}{HTML}{808000}%
\definecolor{H}{HTML}{000000}%
\begin{tikzpicture}
	\footnotesize
	\begin{axis}[%
		width=0.8\textwidth,
		height=0.6\textwidth,
		at={(1.128in,0.894in)},
		scale only axis,
		xmin=0,
		xmax=6,
		xlabel={CNN Network},
		xtick={0, 1, 2, 3, 4, 5, 6},
		xticklabels={$ $, A, B, C, D, E},
		ymax=100000000,
            ymode=log,
		ylabel = {{Proving time (sec) (log)}},
		ylabel shift=-5pt,
		yticklabel shift={0cm},
		axis background/.style={fill=white},
		legend columns=2,
		legend style={legend cell align=left, align=left, fill=none, draw=none,inner sep=-0pt, row sep=0pt, font = \tiny},
		legend pos = north west,
		ymajorgrids,
		xmajorgrids,
		grid style={line width=.5pt, draw=gray!90,dashed},
		major grid style={line width=.2pt,draw=gray!50},
		minor y tick num=5,
            label style={font=\scriptsize},  
            tick label style={font=\scriptsize}  
		]

		\addplot [color=A, solid, mark=diamond*, mark options={solid, A}]
		table[row sep=crcr]{%
			1 265.8 \\ 
			2 410.4\\ 
			3 834\\ 
			4 844.8\\ 
			5 1002\\ 
		};
		\addlegendentry{\sys}
  
		\addplot [color=B, dashed, mark=square*, mark options={solid, B}]
		table[row sep=crcr]{%
			1 27716.6\\ %
			2 30858.93\\ %
			3 35869.68\\ %
			4 46740.47\\ %
			5 69755.96\\ %
		};
		\addlegendentry{\FirstBaseline}

		\addplot [color=teal, dashed, mark=otimes*, mark options={solid, teal}]
		table[row sep=crcr]{%
			1 612515.11\\ %
			2 682225.36\\ %
			3 793351.2\\ %
			4 1034539.08\\ %
			5 1545091.41\\ %
		};
		\addlegendentry{\SecondBaseline}



	\end{axis}

\end{tikzpicture}%
			\end{subfigure}\hspace{4mm}
			\begin{subfigure}{0.3\textwidth}
%
%
\definecolor{A}{HTML}{e6194B}%
\definecolor{B}{HTML}{f58231}%
\definecolor{C}{HTML}{4363d8}%
\definecolor{D}{HTML}{911eb4}%
\definecolor{E}{HTML}{3cb44b}%
\definecolor{F}{rgb}{0.92900,0.69400,0.12500}%
\definecolor{G}{HTML}{808000}%
\definecolor{H}{HTML}{000000}%
\begin{tikzpicture}
	\footnotesize
	\begin{axis}[%
		width=0.8\textwidth,
		height=0.6\textwidth,
		at={(1.128in,0.894in)},
		scale only axis,
		xmin=0,
		xmax=6,
		xlabel={CNN Network},
		xtick={0, 1, 2, 3, 4, 5, 6},
		xticklabels={$ $, A, B, C, D, E},
		ymax=10000,
            ymode=log,
		ylabel = {Verification time (sec) (log)},
		ylabel shift=-5pt,
		yticklabel shift={0cm},
		axis background/.style={fill=white},
		legend columns=2,
		legend style={legend cell align=left, align=left, fill=none, draw=none,inner sep=-0pt, row sep=0pt, font = \tiny},
		legend pos = north west,
		ymajorgrids,
		xmajorgrids,
		grid style={line width=.5pt, draw=gray!90,dashed},
		major grid style={line width=.2pt,draw=gray!50},
		minor y tick num=5,
            label style={font=\scriptsize},  
            tick label style={font=\scriptsize}  
		]

		\addplot [color=A, solid, mark=diamond*, mark options={solid, A}]
		table[row sep=crcr]{%
			1 1.7\\ 
			2 2.04\\
			3 2.51\\ 
			4 2.50\\ 
			5 2.80\\
		};
		\addlegendentry{\sys}
  
		\addplot [color=B, dashed, mark=square*, mark options={solid, B}]
		table[row sep=crcr]{%
			1 28.96\\ %
			2 32.07\\ %
			3 37.02\\ %
			4 47.78\\ %
			5 70.55\\ %
		};
		\addlegendentry{\FirstBaseline}

		\addplot [color=teal, dashed, mark=otimes*, mark options={solid, teal}]
		table[row sep=crcr]{%
			1 608.08\\ %
			2 677.06\\ %
			3 787.02\\ %
			4 1025.69\\ %
			5 1530.90\\ %
		};
		\addlegendentry{\SecondBaseline}

	\end{axis}

\end{tikzpicture}%
			\end{subfigure}\hspace{4mm}
            \begin{subfigure}{0.3\textwidth}
%
%
\definecolor{A}{HTML}{e6194B}%
\definecolor{B}{HTML}{f58231}%
\definecolor{C}{HTML}{4363d8}%
\definecolor{D}{HTML}{911eb4}%
\definecolor{E}{HTML}{3cb44b}%
\definecolor{F}{rgb}{0.92900,0.69400,0.12500}%
\definecolor{G}{HTML}{808000}%
\definecolor{H}{HTML}{000000}%
\begin{tikzpicture}
	\footnotesize
	\begin{axis}[%
		width=0.8\textwidth,
		height=0.6\textwidth,
		at={(1.128in,0.894in)},
		scale only axis,
		xmin=0,
		xmax=6,
		xlabel={CNN Network},
		xtick={0, 1, 2, 3, 4, 5, 6},
		xticklabels={$ $, A, B, C, D, E},
		ymax=1000000,
            ymode=log,
		ylabel = {Proof size (KB) (log)},
		ylabel shift=-5pt,
		yticklabel shift={0cm},
		axis background/.style={fill=white},
		legend columns=2,
		legend style={legend cell align=left, align=left, fill=none, draw=none,inner sep=-0pt, row sep=0pt, font = \tiny},
		legend pos = north west,
		ymajorgrids,
		xmajorgrids,
		grid style={line width=.5pt, draw=gray!90,dashed},
		major grid style={line width=.2pt,draw=gray!50},
		minor y tick num=5,
            label style={font=\scriptsize},  
            tick label style={font=\scriptsize}  
		]

		\addplot [color=A, solid, mark=diamond*, mark options={solid, A}]
		table[row sep=crcr]{%
			1 252\\
			2 326\\
			3 325\\ 
			4 332.1\\ 
			5 367.2\\
		};
		\addlegendentry{\sys}
  
		\addplot [color=B, dashed, mark=square*, mark options={solid, B}]
		table[row sep=crcr]{%
			1 3133.04\\ %
			2 3566.28\\ %
			3 3979.68\\ %
			4 5108.54\\ %
			5 7498.54\\ %
		};
		\addlegendentry{\FirstBaseline}

		\addplot [color=teal, dashed, mark=otimes*, mark options={solid, teal}]
		table[row sep=crcr]{%
			1 63950.12\\ %
			2 71189.41\\ %
			3 82729.62\\ %
			4 107776.54\\ %
			5 160796.47\\ %
		};
		\addlegendentry{\SecondBaseline}



qq	\end{axis}

\end{tikzpicture}%
			\end{subfigure}
		}\vspace{-.2em}
  
	\caption{Performance of \sys~and the baselines on different CNN networks (with increasing \# parameters)}\label{fig:exp:IMS} 
	
\end{figure*}

\begin{figure*}[!t] 
	\centering
		\resizebox{1.01\textwidth}{!}{
			\begin{subfigure}{0.5\textwidth}
				\begin{tikzpicture}
\begin{axis}[
    ybar,
    xlabel={Input size},
    ylabel={{Proving time (sec)}},
    ylabel shift=-6pt,
    ymin=1e-0,  
    ymax=1e10,   
    ymode=log,  
    legend style={at={(0.005,0.99)}, anchor=north west, font = \small, fill=none, draw=none},
    width=1\textwidth,  
    bar width=2 pt,  
    height=0.77\textwidth,    
    symbolic x coords={32$\times$32, 64$\times$64, 128$\times$128, 256$\times$256},  
    xtick=data,  
    nodes near coords={\pgfmathprintnumber{\pgfplotspointmeta}}, 
    nodes near coords align={vertical}, 
    point meta=explicit symbolic, 
    enlarge x limits=0.13,
    label style={font=\large},  
    tick label style={font=\large}  
]

\addplot[fill=magenta] coordinates {
    (32$\times$32, 27.78) [3]
    (64$\times$64, 27.78) [3]
    (128$\times$128, 27.78) [3]
    (256$\times$256, 27.78) [3]
};

\addplot[fill=cyan] coordinates {
    (32$\times$32, 24574.26) [3] %
    (64$\times$64, 97952.05) [3] %
    (128$\times$128, 391463.22) [3] %
    (256$\times$256, 1565507.89) [3] %
};

\addplot[fill=gray] coordinates {
    (32$\times$32, 542636.18) [3] %
    (64$\times$64, 2169924.78) [3] %
    (128$\times$128, 8679079.22) [3] %
    (256$\times$256, 34715696.96) [3] %
};

\addplot[fill=magenta] coordinates {
    (32$\times$32, 100.7) [5]
    (64$\times$64, 100.7) [5]
    (128$\times$128, 100.7) [5]
    (256$\times$256, 100.7) [5]
};

\addplot[fill=cyan] coordinates {
    (32$\times$32, 59830) [5] %
    (64$\times$64, 255164.4) [5] %
    (128$\times$128, 1053487.6) [5] %
    (256$\times$256, 4280751.6) [5] %
};

\addplot[fill=gray] coordinates {
    (32$\times$32, 1324778.65) [5] %
    (64$\times$64, 5657600.88) [5] %
    (128$\times$128, 23365656.93) [5] %
    (256$\times$256, 94951415.44) [5] %
};

\addplot[fill=magenta] coordinates {
    (32$\times$32, 131.71) [7]
    (64$\times$64, 131.71) [7]
    (128$\times$128, 131.71) [7]
    (256$\times$256, 131.71) [7]
};

\addplot[fill=cyan] coordinates {
    (32$\times$32, 102071.06) [7] %
    (64$\times$64, 468280.6) [7] %
    (128$\times$128, 1999702.29) [7] %
    (256$\times$256, 8258556.18) [7] %
};

\addplot[fill=gray] coordinates {
    (32$\times$32, 2261883.71) [7] %
    (64$\times$64, 10385458.50) [7] %
    (128$\times$128, 44356771.26) [7] %
    (256$\times$256, 183196049.49) [7] %
};

\legend{\sys, \FirstBaseline, \SecondBaseline}  

\end{axis}
\end{tikzpicture}
			\end{subfigure}\hspace{1em}
			\begin{subfigure}{0.5\textwidth}
				\begin{tikzpicture}
\begin{axis}[
    ybar,
    xlabel={Input size},
    ylabel={Verification time (sec)},
    ylabel shift=-11pt,
    ymin=1e-1,  
    ymax=1e6,   
    ymode=log,  
    legend style={at={(0.005,0.99)}, anchor=north west, font = \small, fill=none, draw=none},
    width=1\textwidth,  
    bar width=2pt,  
    height=0.77\textwidth,        
    symbolic x coords={32$\times$32, 64$\times$64, 128$\times$128, 256$\times$256},  
    xtick=data,  
    nodes near coords={\pgfmathprintnumber{\pgfplotspointmeta}}, 
    nodes near coords align={vertical}, 
    point meta=explicit symbolic, 
    enlarge x limits=0.13,
    label style={font=\large},  
    tick label style={font=\large}  
]

\addplot[fill=magenta] coordinates {
    (32$\times$32, 0.64) [3]
    (64$\times$64, 0.64) [3]
    (128$\times$128, 0.64) [3]
    (256$\times$256, 0.64) [3]
};

\addplot[fill=cyan] coordinates {
    (32$\times$32, 25.85) [3] %
    (64$\times$64, 98.45) [3] %
    (128$\times$128, 388.87) [3] %
    (256$\times$256, 1550.52) [3] %
};

\addplot[fill=gray] coordinates {
    (32$\times$32, 538.94) [3] %
    (64$\times$64, 2149.18) [3] %
    (128$\times$128, 8590.18) [3] %
    (256$\times$256, 34354.16) [3] %
};

\addplot[fill=magenta] coordinates {
    (32$\times$32, 1.02) [5]
    (64$\times$64, 1.02) [5]
    (128$\times$128, 1.02) [5]
    (256$\times$256, 1.02) [5]
};

\addplot[fill=cyan] coordinates {
    (32$\times$32, 60.73) [5] %
    (64$\times$64, 254.01) [5] %
    (128$\times$128, 1043.91) [5] %
    (256$\times$256, 4237.12) [5] %
};

\addplot[fill=gray] coordinates {
    (32$\times$32, 1312.89) [5] %
    (64$\times$64, 5600.34) [5] %
    (128$\times$128, 23122.97) [5] %
    (256$\times$256, 93959.14) [5] %
};

\addplot[fill=magenta] coordinates {
    (32$\times$32, 1.12) [7]
    (64$\times$64, 1.12) [7]
    (128$\times$128, 1.12) [7]
    (256$\times$256, 1.12) [7]
};

\addplot[fill=cyan] coordinates {
    (32$\times$32, 102.53) [7] %
    (64$\times$64, 464.87) [7] %
    (128$\times$128, 1980.14) [7] %
    (256$\times$256, 8172.96) [7] %
};

\addplot[fill=gray] coordinates {
    (32$\times$32, 2240.19) [7] %
    (64$\times$64, 10278.70) [7] %
    (128$\times$128, 43894.28) [7] %
    (256$\times$256, 181279.75) [7] %
};

\legend{\sys, \FirstBaseline, \SecondBaseline}  

\end{axis}
\end{tikzpicture}
			\end{subfigure}\hspace{1em}
            \begin{subfigure}{0.5\textwidth}
				\begin{tikzpicture}
\begin{axis}[
    ybar,
    xlabel={Input size},
    ylabel={Proof size (KB)},
    ylabel shift=-5pt,    
    ymin=1e1,  
    ymax=1e8,   
    ymode=log,  
    legend style={at={(0.005,0.99)}, anchor=north west, font = \small, fill=none, draw=none},
    width=1\textwidth,  
    bar width=2pt,  
    height=0.77\textwidth,            
    symbolic x coords={32$\times$32, 64$\times$64, 128$\times$128, 256$\times$256},  
    xtick=data,  
    nodes near coords={\pgfmathprintnumber{\pgfplotspointmeta}}, 
    nodes near coords align={vertical}, 
    point meta=explicit symbolic, 
    enlarge x limits=0.13,
    label style={font=\large},  
    tick label style={font=\large}  
]

\addplot[fill=magenta] coordinates {
    (32$\times$32, 127.91) [3]
    (64$\times$64, 127.91) [3]
    (128$\times$128, 127.91) [3]
    (256$\times$256, 127.91) [3]
};

\addplot[fill=cyan] coordinates {
    (32$\times$32, 2806.72) [3] %
    (64$\times$64, 10426.51) [3] %
    (128$\times$128, 40905.67) [3] %
    (256$\times$256, 162822.29) [3] %
};

\addplot[fill=gray] coordinates {
    (32$\times$32, 56693.33) [3] %
    (64$\times$64, 225684.27) [3] %
    (128$\times$128, 901648.03) [3] %
    (256$\times$256, 3605503.07) [3] %
};

\addplot[fill=magenta] coordinates {
    (32$\times$32, 186.98) [5]
    (64$\times$64, 186.98) [5]
    (128$\times$128, 186.98) [5]
    (256$\times$256, 186.98) [5]
};

\addplot[fill=cyan] coordinates {
    (32$\times$32, 6467.8) [5] %
    (64$\times$64, 26751.96) [5] %
    (128$\times$128, 109652.44) [5] %
    (256$\times$256, 444782.04) [5] %
};

\addplot[fill=gray] coordinates {
    (32$\times$32, 137917.39) [5] %
    (64$\times$64, 587873.04) [5] %
    (128$\times$128, 2426822.21) [5] %
    (256$\times$256, 9860872.05) [5] %
};

\addplot[fill=magenta] coordinates {
    (32$\times$32, 195.89) [7]
    (64$\times$64, 195.89) [7]
    (128$\times$128, 195.89) [7]
    (256$\times$256, 195.89) [7]
};

\addplot[fill=cyan] coordinates {
    (32$\times$32, 10854.24) [7] %
    (64$\times$64, 48882.64) [7] %
    (128$\times$128, 207910.45) [7] %
    (256$\times$256, 857850.21) [7] %
};

\addplot[fill=gray] coordinates {
    (32$\times$32, 235234.03) [7] %
    (64$\times$64, 1078852.39) [7] %
    (128$\times$128, 4606710.97) [7] %
    (256$\times$256, 19024915.59) [7] %
};

\legend{\sys, \FirstBaseline, \SecondBaseline}  

\end{axis}
\end{tikzpicture}
			\end{subfigure}
		}\vspace{-.4em}
  
	\caption{Performance of \sys~and the baselines for a single convolution}\label{fig:exp:conv} 
	
\end{figure*}

\PP{Counterpart Comparison}%
\newarman{
To our knowledge, 
we are the first to propose a 
privacy-preserving and verifiable CNN inference scheme that fully preserves the privacy of the client's data sample while ensuring inference verifiability.
Therefore, we compared our scheme to two different baseline settings.
In the first setting, termed \SecondBaseline, we directly hardcode the CNN operations that we aim to prove into the arithmetic circuit.
For the latter, 
named \FirstBaseline,
we use the EC Curve Embedding but lack the probabilistic matrix multiplication check.
This involves 
constructing arithmetic constraints 
for every single computation 
(including EC point multiplications and EC point additions) 
in CNN layers.
In both settings, 
we use the same back-end CP-SNARK scheme 
(i.e., Spartan \cite{spartan}).
}
\newarman{
The most closely related works to ours are zkCNN \cite{zkCNN} and pvCNN \cite{pvcnn}.
However, 
since these schemes offer different functionalities, 
(e.g., lack of full data privacy)
we will only compare our scheme with these works conservatively.
We will also discuss the cost of transforming these schemes to achieve full data privacy.
}

\PP{Evaluation Metrics}%
We benchmark the performance of our scheme and the baseline by measuring their proving time, verification time, proof size\newarman{, and inference accuracy.}

\subsection{Overall Results} \label{sec:OverallResults}%
\autoref{fig:exp:IMS} shows the performance of our proposed scheme in proving time, verification time, and proof size compared to the baseline.
Our $\sys$ scheme is \newarman{1-6} orders of magnitudes more efficient than the baseline in all metrics.
\newarman{
As shown in \autoref{fig:exp:IMS}, our proving time ranges from 266 to 1002 sec for all CNN networks 1 to 5, compared to the proving time 27717 to 69756 sec for the \FirstBaseline, and 612515 to 1545091 sec for the \SecondBaseline. 
Thus, our proving time is  43--104$\times$ faster than the \FirstBaseline~and 951--2304$\times$ faster than the \SecondBaseline. 
}
\newarman{
Our verification time is 15--25$\times$ faster than the \FirstBaseline~and 314-547$\times$ faster than the \SecondBaseline, taking between 1.7 and 2.8 sec compared to 29 and 70.6 sec and 608 and 1531 sec for the \FirstBaseline~and \SecondBaseline, respectively.
}
\newarman{
In our scheme, the proof size is much smaller than the baseline, ranging from 252 to 367 KB compared to 3133 and 7499 KB for the \FirstBaseline~and 63950
160797 KB for the \SecondBaseline. 
Thus, the proof size in $\sys$ is 11--438$\times$ times less than the baseline.}

Next, we focus on evaluating our proposed scheme on the convolutional layer. This is because optimizing the convolutional layer operations is essential in CNN architecture, as it accounts for the majority of computation operations. 
Thus, enhancing the efficiency of the convolutional layer operations can significantly improve the overall performance of our CNN.

\PP{Convolution}%
\autoref{fig:exp:conv} illustrates 
the comparison between our $\sys$ scheme and the baseline within the convolutional layer in terms of proving time, verification time, and proof size.
We conducted this comparison across various input sizes, ranging from 32$\times$32 to 256$\times$256, and filter sizes ranging from 3$\times$3 to 7$\times$7. 
We set the padding and stride to 1 for the 2D convolutional layer throughout the evaluations.
With a 256$\times$256 input and a 3$\times$3 filter, 
the convolutional proving time in our scheme is 27.78 sec, compared to 1565508 sec in the \FirstBaseline~and 34715697 sec in the \SecondBaseline. Thus, our scheme is 56354--1249665$\times$ faster in proving. 
Furthermore, $\sys$ achieves a verification time of only 0.64 sec and a proof size of 128 KB, compared with the verification time of 1551 sec and the proof size of 162822 KB in the \FirstBaseline, 
and 34354 sec and 3605503 KB in the \SecondBaseline.
This represents a remarkable improvement of 
2423--53678$\times$ in verification time and 1273--28188$\times$ in proof size.

As depicted in \autoref{fig:exp:conv}, enlarging the input size with the same filter does not increase proving time, verification time, or proof size in our scheme.
This is because in $\sys$ each depends solely on the filter dimension due to random linear combinations.
Specifically, 
for a convolutional layer with a 7$\times$7 filter,
our scheme 
takes 132 sec to prove,
1.12 sec to verify, 
and produces a proof size of 196 KB 
across input sizes ranging from 
32$\times$32 to 256$\times$256. 
In contrast, 
the baseline takes significantly longer to prove and verify the correctness of larger input sizes as well as the proof size. 
For instance, \FirstBaseline~takes 102071 to 8258556 sec, and \SecondBaseline~takes 2261884 to 183196050 sec to prove a convolutional layer with a 7$\times$7 filter for input sizes ranging from 32$\times$32 to 256$\times$256. 
Similarly, the verification time and proof size range from 103 to 8173 seconds and 10854 to 857850 KB for the \FirstBaseline, and 2240 to 181280 sec and 235234 to 19024916 KB for the \SecondBaseline, respectively.

Within the convolutional layer, the proving and verification times, as well as the proof size in $\sys$, increase only slightly with the filter size, for a fixed input size.
\newarman{
For example, 
for a 32$\times$32 input and 
filter sizes ranging from 3$\times$3 to 7$\times$7, 
our scheme takes between 27.8 and 132 sec 
to generate a proof and 
between 0.64 and 1.12 sec to verify the proof, 
with the proof size ranging between 128 and 196 KB.
}
In contrast, the proving and verification times, as well as the proof size in the baseline, increase significantly with the filter size.
\newarman{
For a 32$\times$32 input and filter sizes ranging from 3$\times$3 to 7$\times$7, 
the \FirstBaseline~takes between 24574 and 102071 sec to generate a proof, between 25.9 and 103 seconds to verify the proof, and the proof size ranges from 2807 to 10854 KB. 
Meanwhile, 
the \SecondBaseline~takes 542636 to 2261884 sec and 539 to 2240 sec to generate and verify a proof, with proof sizes ranging from 56693 to 235234 KB. 
}
\newarman{
This shows that our scheme is 
775--19519$\times$ faster in proving, 
40--2000$\times$ faster in verifying,
and the proof size in our scheme is 22--1201$\times$ less
than that of the baseline 
when the input size is 32$\times$32 and 
the filter sizes range from 3$\times$3 to 7$\times$7.
}
%

\begin{table*}[h]
\small
\centering 
\begin{threeparttable}
\caption{Performance of our $\sys$ scheme on LeNet \cite{lenet}} 
\label{tab:lenet}
\renewcommand{\arraystretch}{0.9} 
\begin{tabularx}{\linewidth}{l *{3}{R} c *{3}{R} c *{3}{R}}\toprule
& \multicolumn{3}{c}{$\sys$} & \phantom{abc}& \multicolumn{3}{c}{\FirstBaseline} &
\phantom{abc} & \multicolumn{3}{c}{\SecondBaseline}\\
\cmidrule{2-4} \cmidrule{6-8} \cmidrule{10-12}
& $\Prover$ (s) & $\Verifier$ (s) & |$\Pi$| (KB) && $\Prover$ (s) & $\Verifier$ (s) & |$\Pi$| (KB) && $\Prover$ (s) & $\Verifier$ (s) & |$\Pi$| (KB)\\ \midrule
$1^{\text{st}}$ Conv. & 417 & 1.8 & 298 && 312225 & 310.5 &  32677 && 6828361 & 6387 & 696674\\
$1^{\text{st}}$ Avg Pool & 34 & 0.6 & 104 && 34 & 0.6 & 104 && 713.3 & 12.1 &2302\\
$2^{\text{nd}}$ Conv. & 1538 & 3.2 & 477 && 398215 & 395.5 &  41607 && 8617373 & 8211& 916186\\
$2^{\text{nd}}$ Avg Pool & 16 & 0.4 & 83 && 16 & 0.4 & 83 &&  327.2& 8.7&1714\\
$3^{\text{rd}}$ Conv. & 8043 & 6.6 & 856 && 127507 & 127.7 &  13496 && 2793806& 2603&298693\\
$1^{\text{st}}$ FC & 401 & 1.8 & 298 && 26867 & 28.1 &  3045 && 570730& 583.3& 65563\\
$2^{\text{nd}}$ FC & 250 & 1.5 & 223 && 2344 & 3.8 &  498 && 49299 & 78.1&10863\\
\midrule
Total & 10699 & 15.9 & 2339 && 867207 & 866.6 &  91510 && 18914609 & 17883 & 1991995\\
\bottomrule 
\end{tabularx}
\end{threeparttable}
\end{table*}

\PP{Experiments with LeNet}%
We compared the performance of our $\sys$ scheme against the baselines on LeNet \cite{lenet} to explore our scheme's performance within a real network setting.
As shown in \autoref{tab:lenet}, our $\sys$ scheme is faster and more efficient at proving, verifying, and proof size. 
\newarman{
Specifically, 
our proving time is 10699 sec 
for 7508 point multiplications and 16864 point additions, 
compared to the \FirstBaseline~and \SecondBaseline, which take days (241 to 5254 hours)
for 652040 point multiplications and 637248 point additions. 
This makes our proving time 81--1768$\times$ faster than the baseline.
}
\newarman{
For verification time, our scheme only takes 15.9 sec, 
while the \FirstBaseline~and \SecondBaseline~take 866.6 and 17883 sec, respectively.
}
\newarman{
Lastly, our proof size is 2339 KB, smaller than \FirstBaseline~and \SecondBaseline, which are 91510 and 1991995 KB respectively. Thus, it is 39--852$\times$ smaller than the baseline.
}

\newarman{
In \autoref{tab:lenet}, the increased overhead in the third convolutional layer of $\sys$ is mainly due to computing 120 feature maps. This substantially impacts the overall overhead of the scheme.
}

\PP{Proving Circuit Size}%
To better understand the
proving times in \autoref{fig:exp:IMS} and \autoref{fig:exp:conv}, we present in \autoref{tab:constraints} the number of arithmetic constraints required for proving $n$ EC point multiplications and $n$ EC point additions within R1CS in both $\sys$ and the baseline.
\newarman{
In \sys~and \FirstBaseline, a \textit{single} EC point multiplication and EC point addition requires 3464 and 10 arithmetic constraints, respectively. 
Meanwhile, \SecondBaseline~requires 76834 and 230 constraints to prove the same operations.
}
\newarman{
Since the proving time scales linearly with the number of R1CS constraints, performing one EC point multiplication consumes more time compared to one EC point addition.
This is because proving a single EC point multiplication with a 128-bit scalar involves iterative calls to $\mathsf{PtAdd}$ and $\mathsf{PtDbl}$ gadgets.
Thus, due to the 128-bit random challenge used in the random linear combination, EC point multiplication is more resource-intensive compared to the highly efficient EC point addition, which only requires 10 constraints in \sys.
}

\newarman{
In \autoref{tab:constraints2},
we compare the total number of arithmetic constraints 
required for proving the LeNet model. 
Specifically, 
with \sys, 
the total number of arithmetic constraints 
needed to prove all EC point multiplications and additions 
for LeNet layers is 26.2 million,
compared to 2265 million and 50245 million in \FirstBaseline~and \SecondBaseline, respectively.
Thus, \sys~requires 86--1918$\times$ fewer constraints to prove the LeNet model compared to the baselines.
}

\begin{table}[t]
\small
\centering 
\caption{Number of R1CS constraints for $n$ EC point multiplications and point additions.}
\begin{tabular}{@{} l@{\hspace{2.5em}}>{\raggedleft\arraybackslash}p{2.8cm}c>{\raggedleft\arraybackslash}p{2.5cm} @{}}    \toprule
Operations & \# Constraints in $\sys$ /\ \FirstBaseline && \# Constraints in \SecondBaseline \\ \midrule
EC point mult & $3464 \cdot {n}$ && $76834 \cdot {n}$ \\ 
EC point add & $10 \cdot {n}$ && $230 \cdot {n}$ \\ 
\bottomrule
\vspace{0.1cm}
\end{tabular}
\label{tab:constraints}
\end{table}

\begin{table}[t]
\small
\centering 
\caption{Number of R1CS constraints in LeNet \cite{lenet}}
\begin{tabular}{@{} l@{\hspace{-0.3em}}>{\raggedleft\arraybackslash}p{2.14cm}c@{\hspace{-1.5em}}>{\raggedleft\arraybackslash}p{2.4cm}c@{\hspace{-0.5em}}>{\raggedleft\arraybackslash}p{2.33cm} @{}}    \toprule
Operations & \# Constraints in $\sys$ && \# Constraints in \FirstBaseline && \# Constraints in \SecondBaseline \\ \midrule
EC point mult & \num{26007712} && \num{2258666560} && \num{50098841360} \\ 
EC point add & \num{168640} && \num{6372480} && \num{146567040} \\ 
\midrule
Total & \num{26176352} && \num{2265039040} && \num{50245408400}\\ 
\bottomrule
\end{tabular}
\label{tab:constraints2}
\end{table}

\PP{Accuracy} \label{sec:accuracy}%
\newarman{
We assess the impact of fixed-point representation on the inference accuracy of the LeNet model on MNIST, as this is what we used in \sys.
In the original non-secure setting with floating-point data and model parameters, the LeNet model achieved 97.87\% accuracy.
When using fixed-point representation and truncation after each layer to keep output within a decryptable range (e.g., $\leq35$ bits) as in \sys, the accuracy slightly dropped to 97.74\%. 
This decrease (e.g., 0.13\%) was expected due to the potential precision loss when converting model parameters and data samples to fixed-point representation and applying truncation.
}

\PP{Experiments with CIFAR-10}%
We further evaluate the performance of \sys~on the LeNet model using CIFAR-10 as the color image dataset. Unlike MNIST which consists of grayscale images with a single channel, CIFAR-10 images have three channels (e.g., RGB). This requires additional parameters in the first convolutional layer, resulting in an increase in the number of point multiplications and point additions from 7508 to 8108 and from 16864 to 36256, respectively, for the CIFAR-10 dataset compared to MNIST. 
Consequently, for the CIFAR-10 dataset, \sys~requires 11951 sec for proving time, 18 sec for verification time, and generates a proof size of 2611 KB. 
In comparison, \FirstBaseline~requires a proving time of 1778111 sec (resp. 37068924 sec for \SecondBaseline), a verification time of 1768.1 sec (resp. 37314 sec for \SecondBaseline), and produces a proof size of 186194 KB (resp. 3902027 KB for \SecondBaseline). 
Thus, \sys~is 71--149$\times$ more efficient than \FirstBaseline~and 1494--3102$\times$ more efficient than \SecondBaseline. 

\PP{Discussion}%
In $\sys$,
we aim to achieve client data privacy protection using AHE,
as well as CNN inference verifiability through CP-SNARK. 
There are two relevant prior works, zkCNN \cite{zkCNN} and pvCNN \cite{pvcnn}, both focusing on CNN inference verifiability with Commit-and-Prove zero-knowledge SNARK (CP-zkSNARK).
However, unlike $\sys$, 
they do not offer full privacy for client data and CNN inference verifiability simultaneously.
Specifically, 
zkCNN \cite{zkCNN} is a zero-knowledge CNN scheme that only protects the privacy of the CNN parameters without preserving the privacy of the client's data. 
The core technique in zkCNN is the new sumcheck protocol for fast Fourier transform (FFT) and doubly efficient interactive proofs for proving the CNN inference, particularly the convolution and matrix multiplication operations within the FC layers.
Although FFT permits fast proving of convolution operations on two plaintext inputs (client sample and server model),
it may be challenging to apply FFT to the context where the client sample is homomorphically encrypted since the HE ciphertexts operate on somewhat unfriendly FFT algebraic group structures \cite{ecfft1, ecfft2}.

pvCNN \cite{pvcnn} is a privacy-preserving CNN inference scheme that offers verifiability and client data privacy.
Although pvCNN uses similar building blocks as $\sys$ including
HE (fully vs. partially in \sys) and (CP-zkSNARK vs. CP-SNARK in \sys), 
unlike \sys,
pvCNN only offers partial privacy for the client sample.
This is because, in pvCNN, the CNN network is divided into two parts: PriorNet and LaterNet.
In PriorNet, the CNN developer (i.e., the server in our case) evaluates a subset of CNN layers on the client ciphertext and generates a zero-knowledge proof for this portion. 
In LaterNet, the encrypted evaluation of PriorNet is decrypted to a third party to continue the inference computation and zero-knowledge proof on the subsequent CNN layers.
\newarman{
As the third party 
learns the evaluation output of PriorNet, 
therefore, 
it can infer client data information. 
Moreover, as the third party has access to a portion of the CNN model of the server (LaterNet), pvCNN only offers partial privacy to the server model (PriorNet privacy).
Meanwhile, in \sys, 
the client data (as well as inference output) is kept encrypted all the time throughout the CNN inference evaluation.
From the performance perspective,
pvCNN relies on Fully Homomorphic Encryption (FHE), which is more costly than AHE.
Thus, we expect that to make pvCNN
achieve the same level of privacy as \sys,  
one has to construct complicated proving circuits for FHE operations and, therefore,
its concrete performance will be significantly slower than \sys.
}

	\section{Related Work}

\PP{Privacy-Preserving ML}%
PPML enables the deployment of machine learning models while preserving the privacy of sensitive information in the training/testing data and model parameters. 
HE \cite{HomomorphicEncryption}, MPC \cite{MPC}, and Differential Privacy \cite{DP1,DP2}, or Trusted Execution Environment (TEE) such as Intel-SGX \cite{SGX} are the most commonly used techniques for PPML.
PPML research covers both the training and inference phases.
A variety of studies focus on PPML in training schemes for machine learning algorithms such as linear regression \cite{ABY, linearReg3}, logistic regression \cite{LRswift, ABY, LR2, LR3, LR4}, Decision Trees (DT) \cite{decisiontree1, decisiontree2}, Support Vector Machines (SVM) \cite{svm1}, Neural Networks (NN) \cite{ABY}, and k-means clustering algorithms \cite{kmean1, kmean2}.
Other research explores PPML techniques for the inference phase (e.g., \cite{LRswift, infer1, infer2, infer3, infer4, infer5, infer6, infer7, infer8, chex}).
Despite significant advancements, a common limitation in \rarman{existing approaches is the lack of a robust} mechanism for guaranteeing computational integrity. 
This limitation can leave vulnerabilities in the integrity of outcomes, especially when clients rely on privacy-preserving computations without the ability to verify the computation's correctness.
\rarman{Moreover, while MPC-based schemes offer potential solutions, they may face limitations in adapting to the MLaaS setting due to their reliance distributed computation among two or more entities.}
SecureDL \cite{secureDL} offers some insights into this challenge with probabilistic verifiability. 
In SecureDL, a client encrypts her pre-trained model and data samples using Leveled Homomorphic Encryption and delegates them to a cloud server for inference computation.
SecureDL achieves verifiability by enabling the client to use some sensitive-data to check whether the server processes encrypted data with the actual encrypted model.
However, this approach does not provide provable verifiability and relies instead on probabilistic methods.

\PP{Verifiable ML}%
\newarman{
Verifiable ML ensures the verifiability of ML algorithm computations by providing provable security. It allows a client to delegate a machine learning task (e.g., training, inference), to a server, and then focus only on verifying the correctness of the training/inference output.
Cryptographic proofs, including Verifiable Computation (VC) protocols \cite{vc1, vc2} and zero-knowledge proofs (ZKPs) \cite{cryptoProof1, cryptoProof2, cryptoProof3, cryptoProof4, spartan, cryptoProof5, bulletproofs} or TEEs \cite{SGX}, are the core techniques for verifiable ML. 
Verifiable ML can be categorized into trusted-setup and trustless schemes. 
Trusted-setup schemes rely on a trusted third party (TTP) to generate a Common Reference String (CRS) such as Gennaro, Gentry, Parno, and Raykova (GGPR) \cite{cryptoProof2}, which introduces potential security vulnerabilities and trust assumptions. 
In contrast, trustless schemes such as Bulletproofs \cite{bulletproofs} and Spartan \cite{spartan} eliminate the need for a TTP or CRS, enhancing security and removing trust assumptions. 
Research in verifiable ML includes various machine learning algorithms like deep neural networks (DNN) inference \cite{DNN1}, linear regression \cite{vML1, vML2}, NN \cite{vML1}, and DT training \cite{vML1}.
Some verifiable ML research incorporates zero-knowledge property
across different machine learning algorithms, including DT \cite{zkml1}, SVM \cite{ezDPS, ezDPS2}, and DNN \cite{zkCNN,zkml2,zkml3, fan2023validating, Mystique}.
Despite their benefits, verifiable ML schemes may not fully preserve the privacy of client data samples during inference, potentially revealing private information from client data.
A closely related work to $\sys$ is pvCNN \cite{pvcnn}, which focuses on privacy preservation within specific CNN layers and ensures inference integrity. However, it lacks full client data privacy, allowing a third party to infer client data.
pvCNN employs trusted-setup ZKPs to generate proofs for targeted CNN inference layers using a CRS.
}

	\section{Conclusion and Future Work}%
%

%
We presented $\sys$, a new privacy-preserving and verifiable CNN scheme allowing clients to securely delegate their CNN inference computations on encrypted data to an MLaaS. 
The server processes the CNN inference for the client's encrypted data sample while providing proof of correct inference computations.
$\sys$ seamlessly integrates HE and VC protocols, ensuring full privacy for client data and computation integrity. 
\sys~makes use of several optimization techniques including curve embedding and probabilistic matrix multiplication to reduce the proving circuit size by a few orders of magnitude.
%
%

%
While \sys~offers inference verifiability and client privacy, there are still significant performance challenges to be addressed to make it more practical. 
Specifically, \sys~requires client-server interaction to handle non-linear functions. 
Furthermore, the proving and verification time in \sys~is still high, especially when applied to complicated models with millions of model parameters.
Finally, an effective CNN inference pipeline generally consists of multiple processing stages (e.g., convolution, activation, pooling, and FC), which may require the server to re-prove the entire inference pipeline in case of system failures during inference.
Therefore,
our future work will seek effective strategies to address the aforementioned challenges.
We will explore techniques that allow the evaluation of non-linear operations directly on the server thereby, mitigating client-server interaction.
To improve the processing delay, we will seek hardware acceleration techniques (e.g., GPU).
Finally, emerging cryptographic techniques such as Incrementally Verifiable Computation \cite{IVC} may be useful for the server to resume proving continuing phases in the inference pipeline in case of system failures.
We leave the exploration of these techniques in our future work.

        \section*{Acknowledgment}
The authors thank the anonymous reviewers for their insightful comments and constructive feedback to improve the quality of this work. 
This work was supported in part by an unrestricted gift from Robert Bosch, 4-VA, and the Commonwealth Cyber Initiative (CCI), an investment in the advancement of cyber R\&D, innovation, and workforce development. For more information about CCI, visit \url{www.cyberinitiative.org}

        \bibliographystyle{IEEEtran}
        \bibliography{PVML}

\appendices
\section{Additional AHE Background} \label{sec:correctnessAndINDCPA}
\newarman{We present the correctness and IND-CPA security of AHE as follows.}

\PP{Correctness} \label{sec:sec:correct}%
An AHE scheme is correct if for any $\secParam, m_1,m_2 \in \mathbb{Z}_n$ s.t. $(\pk,\sk) \gets \HE.\KeyGen(1^\secParam)$, it holds that
 $   \Pr \big[ \HE.\Dec(\sk, \HE.\Enc(\pk,m_1) \boxplus \HE.\Enc(\pk,m_2)) = (m_1 + m_2) \big] = 1 $ and 
   $ \Pr \big[ \HE.\Dec(\sk, {m_2} \boxdot (\HE.\Enc(\pk, m_1))) = m_1 \cdot m_2 \big] = 1$.

\begin{Definition}[IND-CPA Security] \label{def:sec:indcpa}
        Let $\Adv$ be a PPT adversary and $\secParam$ be the security parameter.
        A CPA indistinguishability experiment $\mathsf{Adv}^{\mathsf{cpa}}_{\Adv}(\secParam)$ between the challenger $\Chal$ and $\Adv$ is as follows 
        \begin{enumerate}[label=\arabic*.]
            \item $\Chal$ executes $(\pk,\sk) \gets \HE.\KeyGen(1^\secParam)$ and randomly selects a bit $b \Ra \{0,1\}$. Then, $\Chal$ transmits $\pk$ to $\Adv$.
        
            \item Repeat the following until the adversary decides to stop:
                \begin{enumerate}            
                    \item $\Adv \rightarrow \Chal$: $\Adv$ outputs $m_0$ and $m_1$ of equal length.
        
                    \item $\Chal \rightarrow \Adv$: $\Chal$ executes $\HECipher{C}  \gets \HE.\Enc(\pk,m_b)$.
                \end{enumerate}
            
            \item $\Adv \rightarrow \Chal$: $\Adv$ outputs a bit $b^\prime$.
        \end{enumerate}
        If $b^\prime = b$, $\mathsf{Adv}^{\mathsf{cpa}}_{\Adv}(\secParam) = 1$ ($\Adv$ succeeds); otherwise $\mathsf{Adv}^{\mathsf{cpa}}_{\Adv}(\secParam) = 0$.
        We say $\HE$ is \textit{IND-CPA-secure} if 
        $\Pr \left[
        \mathsf{Adv}^{\mathsf{cpa}}_{\Adv}(\secParam) = 1
        \right] \leq 1/2 + \mathsf{negl(\secParam)}$

\end{Definition}

%
In \autoref{fig:elgamal}, we present exponential ElGamal \cite{exponentialElgamal} encryption scheme on cyclic group \GG.
The decryption in exponential ElGamal makes use of a discrete log solver function ($\mathsf{discretelog}$), which takes $\HECipher{m} \in \GG$ as input, and returns $m$.

\begin{figure}[H]
\centering
\noindent \fbox{\parbox{.99\columnwidth}
	{
		\small
	\underline{$ (\pk,\sk) \gets \HE.\KeyGen(1^\secParam)$:}
        \begin{algorithmic}[1]
	    \State $ x \getsRandom \ZZ_q^*$ where $q$ is the prime order of $\GG$ \label{line:keyGen:x}
            \State $\sk := x$
            \State $\pk := \HECipher{x}$ \\
		\Return $(\pk,\sk)$
	\end{algorithmic}

	\underline{$ \HECipher{c}  \gets \HE.\Enc(\pk,m) $}: 
	
	\begin{algorithmic}[1]
        \setcounter{ALG@line}{4}
	\State $r \getsRandom \ZZ_q^*$
        \State $\HECipher{c_1} = \HECipher{r}$
        \State $\HECipher{c_2} = \HECipher{m} + r\HECipher{x}$
	\State \Return $\HECipher{c} :=(\HECipher{c_1},\HECipher{c_2})$
	\end{algorithmic}

	\underline{$ m \gets \HE.\Dec(\sk,\HECipher{c}) $}: 
	\begin{algorithmic}[1]
        \setcounter{ALG@line}{8}
            \State \parse $\HECipher{c} :=(\HECipher{c_1},\HECipher{c_2})$
            \State $\HECipher{s} = x \scalMult{\HECipher{c_1}}$
	    \State $\HECipher{m} = \HECipher{c_2} + (-1) \scalMult{\HECipher{s}}$            
            \State $m \gets \mathsf{discretelog} (\HECipher{m})$ \Comment{brute-force to find $m$}     
	    \State \Return $m$
	\end{algorithmic}
}}
\caption{Exponential ElGamal Encryption} \label{fig:elgamal}
\end{figure}

\section{Detailed \sys~protocol} \label{sec:appendixDetailedProtocol}
We present the $\relu$ function in \autoref{fig:Relu}, and our detailed scheme protocol in \autoref{fig:PVMLprotocol1}, \autoref{fig:PVMLprotocol2}, and \autoref{fig:PVMLprotocol3}.

\begin{figure}[H]
\centering
\noindent \fbox{\parbox{.99\columnwidth}
	{
        \small
	\underline{$ \HECipher{a'}_2 \gets \relu(\sk, \pk, \HECipher{a}_2,\pp)$:}
        \begin{algorithmic}[1]
        \State ${a} \gets \HE.\Dec(\sk,\HECipher{{a}}_2)$
        \State \textbf{if }{${a} \leq 0$} \textbf{then}
        \State \hspace{2em} $\HECipher{a'}_2 := \HECipher{0}_2$
        \State \textbf{else if }{${a} > 0$} \textbf{then}
        \State \hspace{2em} ${{a}^*} := {a}/2^{\zeta}$ 
        \State \hspace{2em} $\HECipher{a^*}_2 \gets \HE.\Enc(\pk,{a}^*)$
        \State \hspace{2em} $\HECipher{a'}_2 := \HECipher{{{a}}^*}_2$\\
        \Return $\HECipher{a'}_2$
	\end{algorithmic}
 
}}
\caption{$\relu$ function} \label{fig:Relu}
\end{figure}

\begin{figure}[H]
\centering
\noindent \fbox{\parbox{.99\columnwidth}
	{
        \small
	\underline{$(\sk, \pk, \pp) \gets \sys.\Setup(1^\secParam,l)$:}
	\begin{algorithmic}[1] \label{alg:setup}
            \State Find $E_1$ be an EC of a base field $n_1$ and a prime order $q_1$ 
            \State Find $E_2$ be an EC of a base field $q_1$ and a prime order $q_2$
            \State $\pp^\prime$ $\gets$ $\cpzkp.\Setup(1^\secParam)$
            \State $(\pk,\sk) \gets \HE.\KeyGen(1^\secParam)$
            \State $\pp$ $\gets$ $(\pp^\prime, E_1$, $E_2$, $n_1$, $q_1$, $q_2$)\\
		\Return $(\sk, \pk, \pp)$
  
	\end{algorithmic}

        \underline{$\hat{\cm} \gets \sys.\Comm(\mat{W},r,\pp)$}:
        
	\begin{algorithmic}[1] \label{alg:Comm}
        \setcounter{ALG@line}{6}

        \State $ \hat{\cm} \gets \cpzkp.\Comm(\vect{W}, r, \pp) $
	\State \Return $\hat{\cm}$
	\end{algorithmic}

        \underline{$\HECipher{\mat{C}}_2 \gets \sys.\Enc(\pk, \mat{X}, \pp)$}: 

	\begin{algorithmic}[1]
        \setcounter{ALG@line}{8}
        \For{$\text{$i = 1$ to $n$}$} \Comment{$n \times m$: encrypted data sample dimensions}
            \For{$\text{$j = 1$ to $m$}$}
               \State $\HECipher{\mat{C}[i,j]}_2 \gets \HE.\Enc(\pk,\mat{X}[i,j])$
            \EndFor	
        \EndFor	
        \State \Return $\HECipher{\mat{C}}_{2}$
	\end{algorithmic}

}}
\caption{Our proposed \sys~(part 1)} \label{fig:PVMLprotocol1}
\end{figure}

\begin{figure*}[t]
\centering
\noindent \fbox{\parbox{.99\textwidth}
	{\small
        \underline{$(\pi,\vect{t}) \gets \sys.\Infer(\mat{W},\HECipher{\mat{C}}_2,\pp)$}:

	\begin{algorithmic}[1]
        \setcounter{ALG@line}{12}
        \State \parse $\mat{W}:= (\mat{F}, \hat{\mat{W}}, \hat{\vect{b}})$ \Comment{$\mat{F}:$ conv. filter, $\hat{\mat{W}}, \hat{\vect{b}}$: FC layer parameters}
        \State $n^\prime = ((n - k)/\hat{s}) + 1$
        \Comment{$k:$ conv. filter size $\hat{s}:$ stride size}
        \State $m^\prime = ((m - k)/\hat{s}) + 1$
        \For{$\text{$i = 0$ to $n'$}$}
            \For{$\text{$j = 0$ to $m'$}$}
                \State $\HECipher{\mat{A}[i,j]}_2 = \sum_{i'=0}^{k-1} \sum_{j'=0}^{k-1} \mat{F}[i',j'] \scalMult{\HECipher{\mat{C}[i \cdot \hat{s} +i',j \cdot \hat{s}+j']}_2}$        
            \EndFor
        \EndFor
        
        \State Server interacts with the client to compute $\HECipher{\mat{A}^\prime[i,j]}_2 \gets \relu(\sk, \pk, \HECipher{\mat{A}[i,j]}_2,\pp)$
        for each $i \in [0,n^\prime]$ and $j \in [0,m^\prime]$.

        \For{$\text{$i = 0$ to $n^\prime$}$} 
            \For{$\text{$j = 0$ to $m^\prime$}$} 
                \State $\HECipher{\mat{B}[i,j]}_2 = \frac{1}{\hat{k}^2} \sum_{i'=0}^{\hat{k}-1} \sum_{j'=0}^{\hat{k}-1} \HECipher{\mat{A}^\prime[i \cdot \hat{k} + i', j \cdot \hat{k} + j']}_2$
                \Comment{$\hat{k}:$ pooling window size} 
        
            \EndFor
        \EndFor

        \State $\hat{n} = n' / \hat{k}$
        \State $\hat{m} = m' / \hat{k}$       
        \For{$i = 0$ to $\hat{n}$}
            \For{$j = 0$ to $\hat{m}$}
                \State $\HECipher{\vect{d}[i\cdot \hat{m} + j]}_2 = \HECipher{\mat{B}[i,j]}_2$
            \EndFor
        \EndFor     
        
        \For{$\text{$i = 1$ to $h$}$} \Comment{$h:$ FC layer output size}
            \State $\HECipher{\vect{b}[i]}_2 \gets \HE.\Enc(\pk,\vect{\hat{b}}[i])$
        \EndFor
        
        \For{$\text{$i = 1$ to $h$}$}
            \State $\HECipher{\vect{t}[i]}_2 = (\sum_{j=0}^{g} \hat{\mat{W}}^T[i,j] \scalMult{\HECipher{\vect{d}[j]}_2}) + \HECipher{\vect{b}[i]}_2$
            \Comment{$g:$ FC layer input size}
        \EndFor

        \State To compute the next FC layers, the server interacts with the client to compute $\HECipher{\vect{t}^\prime[i]}_2 \gets \relu(\sk, \pk, \HECipher{\vect{t}[i]}_2,\pp)$ for each $i \in [0,h]$. 
        The server repeats the process in lines 29 to 32 with the new weight $\hat{\mat{W}}_1$ and bias $\vect{\hat{b}}_1$ parameters.
        
        \State Let $\phi_{\mathsf{cnv}},\phi_{\mathsf{pl}},\phi_{\mathsf{fc}}$ be arithmetic circuits for convolution, pooling, and FC layer, respectively. The server uses $\mathsf{PtMul}$ and $\mathsf{PtAdd}$ gadgets to generate constraints for
        $\phi_{\mathsf{cnv}},\phi_{\mathsf{fc}}$, and $\mathsf{PtAdd}$ to generate constraints for $\phi_{\mathsf{pl}}$.
        Let $\aux_{\mathsf{cnv}},\aux_{\mathsf{pl}},\aux_{\mathsf{fc}}$ be auxiliary witnesses when creating arithmetic constraints for $\phi_{\mathsf{cnv}},\phi_{\mathsf{pl}},\phi_{\mathsf{fc}}$, respectively.
        The server commits to all auxiliary witnesses 
        as $\cm^\prime \gets \cpzkp.\Comm(\aux, r_2, \pp)$, 
        where $\aux = (\aux_{\mathsf{cnv}},\aux_{\mathsf{pl}},\aux_{\mathsf{fc}})$ and 
        $r_2 \getsRandom {\ZZ_q^*}$, and 
        creates a proof $\pi \gets \cpzkp.\prove(\phi, (\mat{F}, \vect{\hat{\mat{W}}}, \vect{\hat{b}}, \aux), \pp)$, 
        where $\phi$ is the merge arithmetic circuit consisting of $\phi_{\mathsf{cnv}},\phi_{\mathsf{pl}}, \text{and } \phi_{\mathsf{fc}}$.
        
        \State \Return $(\hat{\pi}, \HECipher{\vect{t}}_2)$ where $\hat{\pi} := (\pi, \cm^\prime)$

        \end{algorithmic}

}}
\caption{Our proposed \sys~(part 2)} \label{fig:PVMLprotocol2}
\end{figure*}

\begin{figure}
\centering
\noindent \fbox{\parbox{.99\columnwidth}
	{
        \small
        \underline{$\{\vect{y},\perp\} \gets \sys.\Dec(\sk, \HECipher{\vect{t}}_2,\cm, \hat{\pi}, \pp)$:}
        
	\begin{algorithmic}[1]
        \setcounter{ALG@line}{34}
        \State \parse $\hat{\pi} := (\pi, \cm^\prime)$
        \State $ b \gets \cpzkp.\ver(\pi, \cm, \vect{t}, \pp)$, where $\cm = (\hat{\cm}, \cm^\prime)$. 
                 
        \If {$b=1$}
            \For{$\text{$i = 1$ to $h$}$}
                \State $\vect{y}[i] \gets \HE.\Dec(\sk, \HECipher{\vect{t}[i]}_2)$ 
            \EndFor
        \State \Return $\vect{y}$
        \Else{ \Return $\perp$}
        \EndIf


	\end{algorithmic}


}}
\caption{Our proposed \sys~(part 3)} \label{fig:PVMLprotocol3}
\end{figure}

\begin{figure}[tp]
\centering
\noindent \fbox{\parbox{.99\columnwidth}
	{
		\small
        \PP{Simulator 1 (Simulation of $\sys$ for data sample and inference privacy)} 
        Let $(\sk,\pk,\pp) \gets \sys.\Setup(1^\secParam,l)$ for any $\secParam$ and $l$. 
        \newarman{
        Given $s = n \times m$ as the data sample size,
        $\mat{W}:= (\mat{F}, \hat{\mat{W}}, \hat{\vect{b}})$
        as the server model, 
        where $\mat{F}$ is the convolutional filter and
        $\hat{\mat{W}}, \hat{\vect{b}}$ are FC layer's parameters.
        Let 
        $\hat{s}$ be the stride size and
        $k$, $\hat{k}$ 
        be the convolutional filter and pooling window size.
        Let $g, h$ be the FC layer input and output sizes.
        Given
        $n^\prime = ((n - k)/\hat{s}) + 1$ and 
        $m^\prime = ((m - k)/\hat{s}) + 1$ 
        , along with 
        $\hat{n} = n' / \hat{k}$ and
        $\hat{m} = m' / \hat{k}$,}
        the simulator $\hat{\Sim}$ maintains a copy of the \HE~scheme and acts as follows.

        \begin{itemize}
            
            \item \newarman{$(\HECipher{\vect{t}}_2, \HECipher{\mat{C}}_2) \gets \hat{\Sim}(\pk, s, \pp)$:}
            \newarman{
            $\hat{\Sim}$ executes 
            $\HECipher{c_{i}}_2 \gets \HE.\Enc(\pk,0)$
            for each $i \in [s]$
            and invokes 
            $\HECipher{\mat{A}[i,j]}_2 = \sum_{i'=0}^{k-1} \sum_{j'=0}^{k-1} \mat{F}[i',j'] \scalMult{\HECipher{\mat{C}[i \cdot \hat{s} +i',j \cdot \hat{s}+j']}_2}$
            for $i \in [0, n']$ and $j \in [0,m']$,
            where $\HECipher{\mat{C}}_2 = (\HECipher{c_{1}}_2,\dots,\HECipher{c_s}_2)$. 
            Subsequently, $\hat{\Sim}$ executes  $\HECipher{c'_{i}}_2 \gets \HE.\Enc(\pk,0)$
            for each $i \in [s]$
            and performs
            $\HECipher{\mat{B}[i,j]}_2 = \frac{1}{\hat{k}^2} \sum_{i'=0}^{\hat{k}-1} \sum_{j'=0}^{\hat{k}-1} \HECipher{\mat{C}^\prime[i \cdot \hat{k} + i', j \cdot \hat{k} + j']}_2$,
            for $i \in [0, n']$ and $j \in [0,m']$
            where $\HECipher{\mat{C}'}_2 = (\HECipher{{c'}_{1}}_2,\dots,\HECipher{{c'}_s}_2)$.
            Next, $\hat{\Sim}$ invokes 
            $\HECipher{\vect{d}[i\cdot \hat{n} + j]}_2 = \HECipher{\mat{B}[i,j]}_2$ for $i \in [0, \hat{n}]$ and $j \in [0,\hat{m}]$
            and computes
            $\HECipher{\vect{t}[i]}_2 = (\sum_{j=0}^{g} \hat{\mat{W}}^T[i,j] \scalMult{\HECipher{\vect{d}[j]}_2}) + \HECipher{\vect{b}[i]}_2$
            for $i \in [0,h]$
            Finally, 
            $\hat{\Sim}$ outputs $\HECipher{\mat{C}}_2$ and $\HECipher{\vect{t}}_2$.  
            }
            
        \end{itemize}
 
}}
\caption{Simulator of $\sys$ for data sample and inference privacy.} \label{fig:simulator2}
\end{figure}

\section{Proof of \autoref{thm:Theorem1}}  \label{sec:securityProof}
%
\newarman{We prove the completeness, soundness, and sample and inference privacy of \sys~as follows.}

\PP{Completeness}%
In our $\sys$ scheme (\autoref{fig:PVMLprotocol1}, \autoref{fig:PVMLprotocol2}, and \autoref{fig:PVMLprotocol3}), 
if the inference result $\mat{t}$ and its decryption $\vect{y}$ are correct, 
then $\sys.\Dec$ will not output $\perp$. 
The correctness of $\mat{t}$ within our $\sys$ scheme holds due to the completeness property of the back-end CP-SNARK scheme, 
and the correctness of $\vect{y}$ holds due to the correctness and homomorphic properties of the back-end AHE scheme.

\PP{Soundness}%
Let $\mat{W}$ be the model parameters,
$r$ be the randomness,
$\mat{X}$ be the client data sample such that
$\mat{C} \gets \sys.\Enc(\pk, \mat{X}, \pp)$, $(\pi,\vect{t}) \gets \sys.\Infer(\mat{W},\mat{C},\pp)$, 
and $\cm = \sys.\Comm(\mat{W}, r, \pp)$.
Let $\hat{\mat{W}} \ne \mat{W}$ be an arbitrary model and $\hat{\mat{C}} \ne \mat{C}$ be an arbitrary encrypted sample.
Let $\cm^* \gets \sys.\Comm(\hat{\mat{W}}, r, \pp)$ and $ (\pi^*,\vect{t}^*) \gets \sys.\Infer(\hat{\mat{W}},\mat{C},\pp) $.
Let $\idealFunc'$ be the ideal inference functionality that returns the correct inference result on the input $\mat{W}$ and $\mat{X}$. 
If $\sys.\Dec(\sk, \vect{t}^*, \cm, \pi^*, \pp) \ne \perp$ however $\idealFunc(\mat{W},\mat{C}) \ne \vect{t}^*$ and/or $\sys.\Dec(\sk, \vect{t}^*, \cm, \pi^*, \pp) \ne \vect{y}$ but $\idealFunc^\prime(\mat{W},\mat{X}) = \vect{y}$, then there are two scenarios:
\begin{itemize}
    \item Scenario 1:
        The client accepts the inference proof produced by a dishonest server on the arbitrary model $\hat{\mat{W}}$ and/or the arbitrary sample $\hat{\mat{C}}$.
        %
        $\pi^*$ is verified within the $\sys.\Dec$ algorithm as $\{0,1\} \gets \cpzkp.\ver(\pi^*, \cm, \newarman{\vect{t}^*}, \pp)$. 
        However, the probability of $\cpzkp.\ver(\pi^*, \cm, \newarman{\vect{t}^*}, \pp) = 1$ is negligible in $\secParam$, due to the knowledge soundness of the back-end CP-SNARK, which ensures that if a dishonest prover $\Prover^*$ produces a commitment $\cm$, a proof $\pi^*$ \newarman{and inference result $\vect{t}^*$}, $\cpzkp.\ver(\pi^*, \cm, \newarman{\vect{t}^*}, \pp) = 1$ iff $\cm = \cm^*$, $\pi = \pi^*$, $\mat{W} = \hat{\mat{W}}$, and $\mat{C} = \hat{\mat{C}}$. 
        However, the probability that $\cm^* = \cm$ given $\mat{\hat{W}} \ne \mat{W}$ is negligible in $\secParam$ due to the binding property of the commitment scheme.
        Therefore, it is negligible in $\secParam$ that $\sys.\Dec(\sk, \vect{t}^*, \cm, \pi^*, \pp) \ne \perp$.
        
    \item Scenario 2:   
        Client accepts $\vect{t}^*$ s.t.  $\sys.\Dec(\sk, \vect{t}^*, \cm, \pi^*, \pp) = \vect{y}^*$ and $\vect{y}^* \ne \idealFunc^\prime(\mat{W},\mat{X})$.
        We say that this happens with a negligible probability in $\secParam$ due to the correctness and homomorphic properties of AHE, which ensure that the (linear) inference computation can be evaluated over the ciphertext domain.
        \sys~follows the client-aid model, where the server occasionally sends the encrypted evaluation to the client to perform either non-linear operations (e.g., activation function) or truncation. 
        In \sys, 
        the server always performs homomorphic computation on AHE ciphertext of messages of up to 35 bits
        and 
        the client always receives the encrypted evaluation results of up to 35 bits.
        This ensures that the client can always perform the decryption correctly via the discrete log solver. 
        %
        %
        %
        
        %
        %
        
\end{itemize}

\PP{Data sample and inference privacy}%
%
We construct simulator $ \hat{\Sim} $ for $\sys$ (\autoref{fig:PVMLprotocol1}, \autoref{fig:PVMLprotocol2}, and \autoref{fig:PVMLprotocol3}), shown in \autoref{fig:simulator2}. 
We create a hybrid game consisting of $\textbf{Hybrid}$ $H = (H_1,H_2)$ to prove their indistinguishability as follows
\begin{itemize}
    \item $H_1$: It is the real protocol on the client's encrypted data sample.
    \item $H_2$: It is the simulator $\hat{\Sim}$ in \autoref{fig:simulator2}.
    
\end{itemize}
The server cannot distinguish between $H_1$ and $H_2$ due to the IND-CPA security of  the back-end AHE scheme where the encrypted data sample $\mat{C}$ and encrypted inference result $\vect{t}$ do not reveal any information about the underlying plaintext.

\end{document}